\documentclass[aps,prd,amsmath,amssymb,nofootinbib,showkeys,showpacs,superscriptaddress,twocolumn,10pt]{revtex4-1}

\pdfoutput=1

\usepackage{graphicx}
\usepackage{dcolumn}
\usepackage{bm}
\usepackage{amssymb}
\usepackage{latexsym}
\usepackage{booktabs}
\usepackage{amsmath}
\usepackage{multirow}
\usepackage{enumerate}
\usepackage{url}
\usepackage{subfigure}

\usepackage{float}
\usepackage[colorlinks=true, linkcolor=red, citecolor=blue]{hyperref}

\begin{document}

\title{Taiji-TianQin-LISA network: Precisely measuring the Hubble constant using both bright and dark sirens}

\author{Shang-Jie Jin}
\affiliation{Key Laboratory of Cosmology and Astrophysics (Liaoning) \& College of Sciences, Northeastern University, Shenyang 110819, China}
\author{Ye-Zhu Zhang}
\affiliation{Key Laboratory of Cosmology and Astrophysics (Liaoning) \& College of Sciences, Northeastern University, Shenyang 110819, China}
\author{Ji-Yu Song}
\affiliation{Key Laboratory of Cosmology and Astrophysics (Liaoning) \& College of Sciences, Northeastern University, Shenyang 110819, China}
\author{Jing-Fei Zhang}
\affiliation{Key Laboratory of Cosmology and Astrophysics (Liaoning) \& College of Sciences, Northeastern University, Shenyang 110819, China}
\author{Xin Zhang}\thanks{Corresponding author}\email{zhangxin@mail.neu.edu.cn}
\affiliation{Key Laboratory of Cosmology and Astrophysics (Liaoning) \& College of Sciences, Northeastern University, Shenyang 110819, China}
\affiliation{Key Laboratory of Data Analytics and Optimization for Smart Industry (Ministry of Education), Northeastern University, Shenyang 110819, China}
\affiliation{National Frontiers Science Center for Industrial Intelligence and Systems Optimization, Northeastern University, Shenyang 110819, China}

\begin{abstract}

In the coming decades, the space-based gravitational-wave (GW) detectors such as Taiji, TianQin, and LISA are expected to form a network capable of detecting millihertz GWs emitted by the mergers of massive black hole binaries (MBHBs). In this work, we investigate the potential of GW standard sirens from the Taiji-TianQin-LISA network in constraining cosmological parameters. For the optimistic scenario in which electromagnetic (EM) counterparts can be detected, we predict the number of detectable bright sirens based on three different MBHB population models, i.e., pop III, Q3d, and Q3nod. Our results show that the Taiji-TianQin-LISA network alone could achieve a constraint precision of $0.9\%$ for the Hubble constant, meeting the standard of precision cosmology. Moreover, the Taiji-TianQin-LISA network could effectively break the cosmological parameter degeneracies generated by the CMB data, particularly in the dynamical dark energy models. When combined with the CMB data, the joint CMB+Taiji-TianQin-LISA data offer $\sigma(w)=0.036$ in the $w$CDM model, which is close to the latest constraint result obtained from the CMB+SN data. We also consider a conservative scenario in which EM counterparts are not available. Due to the precise sky localizations of MBHBs by the Taiji-TianQin-LISA network, the constraint precision of the Hubble constant is expected to reach $1.2\%$. In conclusion, the GW standard sirens from the Taiji-TianQin-LISA network will play a critical role in helping solve the Hubble tension and shedding light on the nature of dark energy.

\end{abstract}


\maketitle

\section{Introduction}\label{sec:intro}
The precise measurements of the cosmic microwave background (CMB) anisotropies have ushered in the era of precision cosmology \cite{WMAP:2003elm,WMAP:2003ivt}. Nevertheless, in recent years, with the improvements of the measurement precisions of cosmological parameters, some puzzling tensions appeared. In particular, the tension between the Hubble constant values inferred from the $Planck$ CMB observation (a 0.8\% measurement, assuming the $\Lambda$CDM model) \cite{Planck:2018vyg} and obtained through the distance ladder method (a 1.4\% measurement) \cite{Riess:2021jrx} is now at the $5\sigma$ level. The Hubble tension is now commonly considered a severe crisis for cosmology \cite{Verde:2019ivm,Riess:2019qba}, which has been widely discussed in the literature \cite{Kamionkowski:2022pkx,Riess:2019qba,Verde:2019ivm,Li:2009bn,Zhang:2014dxk,Zhang:2014nta,Zhao:2017urm,Feng:2017nss,Li:2010xjz,Zhang:2014ifa,Guo:2018ans,Perivolaropoulos:2021jda,Gao:2021xnk,DiValentino:2021izs,Abdalla:2022yfr,Cai:2021wgv,Yang:2018euj,DiValentino:2020zio,DiValentino:2019ffd,DiValentino:2019jae,Liu:2019awo,Zhang:2019cww,Ding:2019mmw,Li:2020tds,Wang:2021kxc,Vagnozzi:2021tjv,Vagnozzi:2021gjh,Vagnozzi:2019ezj,Guo:2019dui,Feng:2019jqa,Lin:2020jcb,Gao:2022ahg,Zhao:2022yiv}. The Hubble tension may herald the possibility of new physics beyond the standard model of cosmology.
However, no consensus has been reached on a valid extended cosmological model that can truly solve the Hubble tension. On the other hand, it is crucial to develop cosmological probes that can independently measure the Hubble constant and make an arbitration for the Hubble tension. The gravitational-wave (GW) standard siren method is one of the most promising methods.

The absolute distance information of a GW source could be obtained by analyzing the GW waveform, which is called a standard siren \cite{Schutz:1986gp, Holz:2005df}. If the redshift information could also be obtained from the associated electromagnetic (EM) counterparts (we usually refer to this kind of standard sirens as ``bright sirens''), the established distance-redshift relation can then be used to explore the expansion history of the universe. The only multi-messenger observation event GW170817 has given the first measurement of $H_0$ using the standard siren method, with about 14\% precision \cite{LIGOScientific:2017adf}. The constraint precision of $H_0$ is expected to be 2\% using 50 similar standard siren events \cite{Chen:2017rfc}.
Although for the current detected stellar-mass binary black hole mergers their EM counterparts cannot be detected, the statistical method could be applied in inferring the redshift information for the cosmological parameter estimation (we usually refer to this kind of standard sirens as ``dark sirens''; see e.g. Refs.~\cite{Muttoni:2023prw,Gray:2019ksv,Finke:2021aom,Gair:2022zsa,Song:2022siz,Leandro:2021qlc,DelPozzo:2011vcw,LIGOScientific:2018gmd,Chen:2017rfc,Yang:2022iwn,DES:2019ccw,Nair:2018ign,DES:2020nay,LIGOScientific:2019zcs,Yu:2020vyy,Palmese:2021mjm,LIGOScientific:2021aug} for related works). {The latest result for constraining the Hubble constant using dark sirens comes from the LIGO-Virgo-KAGRA observations, with a precision of $19\%$ \cite{LIGOScientific:2021aug}. However, this result is also far from resolving the Hubble tension.} Recently, GW standard sirens have been widely discussed in the literature \cite{Dalal:2006qt,Cutler:2009qv,Nissanke:2009kt,Cai:2017cbj,Bian:2021ini,Cai:2016sby,Cai:2017aea,Cai:2017plb,Zhang:2019ylr,Chen:2020dyt,Zhao:2010sz,Du:2018tia,Yang:2019bpr,Bachega:2019fki,Chang:2019xcb,He:2019dhl,Zhao:2019gyk,Wang:2019tto,Qi:2021iic,Jin:2021pcv,Zhu:2022dfq,deSouza:2021xtg,Guo:2022,Wu:2022dgy,Jin:2022tdf,Hou:2022rvk,Califano:2022syd,Wang:2022oou,Dhani:2022ulg,Colgain:2022xcz,Cao:2021zpf,Fu:2021huc,Ye:2021klk,Chen:2020zoq,Jin:2022qnj,Mitra:2020vzq,Hogg:2020ktc,Nunes:2020rmr,Jin:2023zhi,Borhanian:2020vyr,Han:2023exn,Li:2023gtu,Jin:2023tou,Jin:2020hmc,Yu:2021nvx,Belgacem:2019tbw,Safarzadeh:2019pis,Zhang:2019ple,Zhang:2018byx,Zhang:2019loq}.

The GW detectors will be greatly developed in the next decades. For the detection of GWs in the frequency band of several hundred (up to several thousand) hertzs, we will have the third-generation ground-based GW detectors, e.g., the Cosmic Explorer \cite{LIGOScientific:2016wof} and the Einstein Telescope \cite{Punturo:2010zz}, which are one order of magnitude more sensitive than the current GW detectors. They are expected to detect standard sirens with redshifts mainly distributed at $z<3$ \cite{Evans:2021gyd}. The cosmological parameter estimations using the future standard sirens detected by the third-generation GW detectors have been forecasted in, e.g. Refs.~\cite{Jin:2020hmc,Yu:2021nvx,Belgacem:2019tbw,Safarzadeh:2019pis,Zhang:2019ple,Zhang:2018byx,Zhang:2019loq}.

{GW standard sirens are expected to play a pivotal role in future cosmological research. However, alongside GW standard sirens, there are several other highly promising cosmological tools, such as fast radio bursts and 21 cm intensity mapping. Recent discussion on cosmological parameter estimations using these promising tools can be found in, e.g., Refs.~\cite{Li:2017mek,Walters:2017afr,SKA:2018ckk,Dai:2023,Zhang:2021yof,Liu:2019jka,Wu:2021vfz,Zhang:2023gye,Zhao:2022bpd,Wu:2022jkf,Chen:2022,Qiu:2021cww,Zhang:2021yof,Zhao:2020ole,Gao:2022ifq}.}

The space-based GW detectors, Taiji \cite{Guo:2018npi,Wu:2018clg,Hu:2017mde}, TianQin \cite{Liu:2020eko,Wang:2019ryf,Luo:2015ght,Luo:2020bls,Milyukov:2020kyg,TianQin:2020hid}, and LISA \cite{Robson:2019,LISA:2017pwj,LISACosmologyWorkingGroup:2022jok}, will open a window for detecting millihertz GWs emitted from the mergers of massive black hole binaries (MBHBs), extreme mass ratio inspirals, etc. According to recent forecasts, the mergers of MBHBs could emit EM radiations in both the radio and optical bands \cite{Palenzuela:2010nf,OShaughnessy:2011nwl,Moesta:2011bn,Kaplan:2011mz,Shi:2011us,Blandford:1977ds,Meier:2000wk,Dotti:2011um}, which are expected to be observed by EM detectors, e.g., Square Kilometer Array (SKA) \cite{SKA-web}, the Extremely Large Telescope (ELT) \cite{ELT-web}, the Large Synoptic Survey Telescope (LSST) \cite{LSST-web}, etc. Compared to the single GW observatory, the GW detection network consisting of multiple GW observatories could improve the measurement precisions of GW parameters \cite{Cai:2023ywp,Wang:2020dkc,Ruan:2019tje,Ruan:2020smc,Zhu:2021bpp,Zhu:2021aat,Lyu:2023ctt,Torres-Orjuela:2023hfd,Wang:2020fwa,Wang:2021njt,Wang:2021srv}. Recently, Ruan \emph{et al.} \cite{Ruan:2020smc} proposed that Taiji and LISA could form a space-based detector network. Compared to the single Taiji observatory, the Taiji-LISA network could greatly improve the ability to locate the GW sources, promoting the applications of the standard siren method in cosmological parameter estimations (related works could refer to, e.g., Refs.~\cite{Wang:2020dkc,Wang:2021srv}).

In the coming decades, Taiji, TianQin, and LISA are expected to observe simultaneously and provided unprecedented high-accuracy localizations of MBHBs. Notably, studies on the cosmological parameter estimation using the standard sirens from the space-based Taiji-TianQin-LISA detector network are still absent. The aim of this work is to explore the potential of the Taiji-TianQin-LISA network in answering three key questions: (i) how well the Taiji-TianQin-LISA network can locate MBHBs, (ii) what precision the cosmological parameters can be measured by the bright sirens from the Taiji-TianQin-LISA network, and (iii) what precision the Hubble constant can be measured using the dark sirens from the Taiji-TianQin-LISA network. Through this work, we hope to deepen our understanding of the standard sirens from the Taiji-TianQin-LISA network for solving the Hubble tension and investigating the potential of measuring dark energy.

This work is organized as follows. In Sec.~\ref{sec:method}, we introduce the methodology used in this work. In Sec.~\ref{sec:simulation}, we detailedly introduce the simulations of bright and dark sirens. In Sec.~\ref{sec:Result}, we give the constraint results and make some relevant discussions. The conclusion is given in Sec.~\ref{sec:con}.

\section{Methodology}\label{sec:method}

\subsection{Cosmological models}
The luminosity distance of a source at redshift $z$ can be written as
\begin{equation}
d_{\rm L} (z)= c(1 + z)\int_0^z {\frac{{dz'}}{{H(z')}}},
\label{equa:dl}
\end{equation}
where $c$ is the speed of light in vacuum, $H(z)$ is the Hubble parameter which describes the expansion rate of the universe at redshift $z$.

In this work, we consider the $\Lambda$CDM, $w$CDM, and $w_0w_a$CDM models to make cosmological analysis.
\begin{itemize}
\item $\Lambda$CDM: the standard model of cosmology with the equation of state (EoS) parameter of dark energy $w=-1$. The form of $H(z)$ is given by
\begin{equation}
    H(z)=H_0\sqrt{\Omega_{\rm m}(1+z)^3+1-\Omega_{\rm m}},
\end{equation}
where $\Omega_{\rm m}$ is the current matter density parameter and $H_0$ is the Hubble constant.

\item $w$CDM: the simplest case for dynamical dark energy model with  $w=\rm constant$. The form of $H(z)$ is given by
\begin{equation}
    H(z)=H_0\sqrt{\Omega_{\rm m}(1+z)^3+(1-\Omega_{\rm m})(1+z)^{3(1+w)}}.
\end{equation}

\item $w_0w_a$CDM: a phenomenological model to explore the evolution of $w$. The forms of $w(z)$ and $H(z)$ are given by
\begin{equation}
    w(z)=w_0+w_a\frac{z}{1+z},
\end{equation}
\begin{equation}
\resizebox{0.9\linewidth}{!}{$H(z)=H_0\sqrt{\Omega_{\rm m}(1+z)^3+(1-\Omega_{\rm m})(1+z)^{3(1+w_0+w_a)}\exp(-\frac{3w_az}{1+z})}$.}
\end{equation}

\end{itemize}

\begin{figure}[!htbp]
\includegraphics[width=0.45\textwidth]{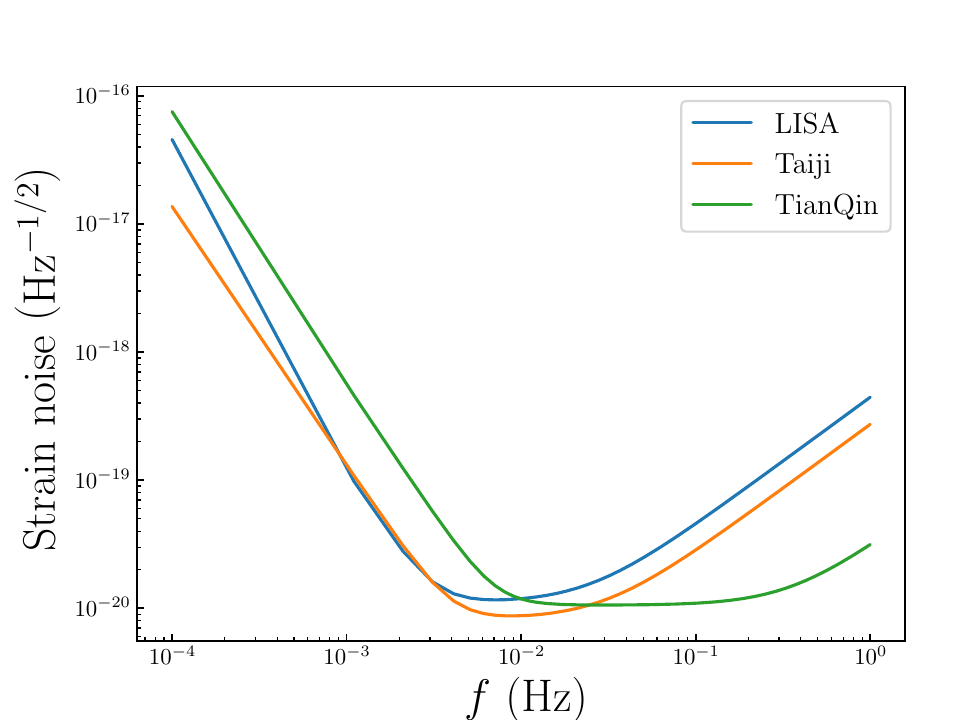}
\centering
\caption{The sensitivity curves of Taiji, TianQin, and LISA. {More details about these sensitivity curves can be found in Refs.~\cite{Guo:2018npi,Wang:2019ryf,LISACosmologyWorkingGroup:2019mwx}. }}\label{fig1}
\end{figure}

Throughout this paper, we adopt the flat $\Lambda$CDM model as the fiducial model to generate the mock standard siren data with $\Omega_{\rm m}=0.3166$ and $H_0=67.27$ ${\rm km\ s^{-1}\ Mpc^{-1}}$ from the constraint results by the $Planck$ 2018 TT,TE,EE+lowE data \cite{Planck:2018vyg}.

\subsection{GW standard sirens}
\begin{figure}[!htbp]
\begin{center}
\includegraphics[width=\linewidth]{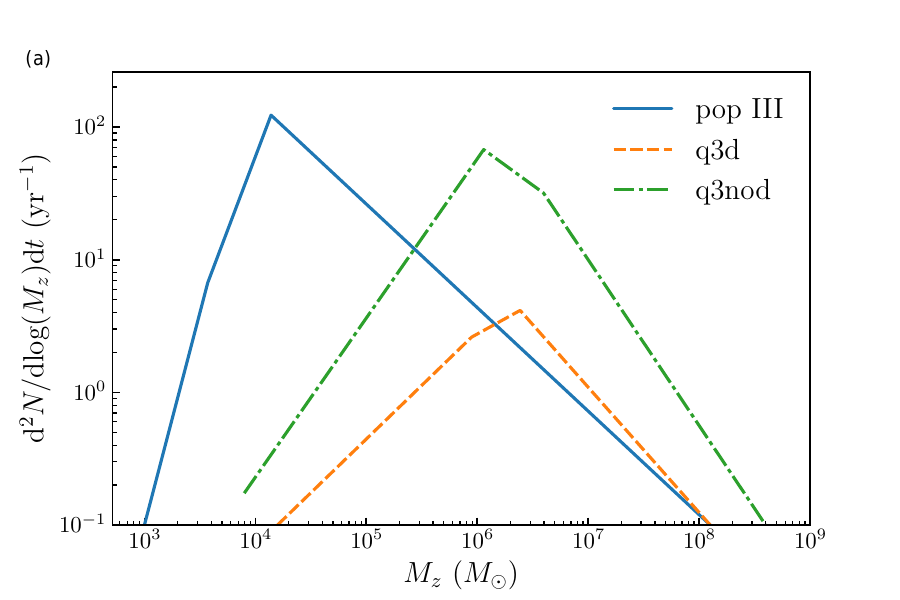}
\includegraphics[width=\linewidth]{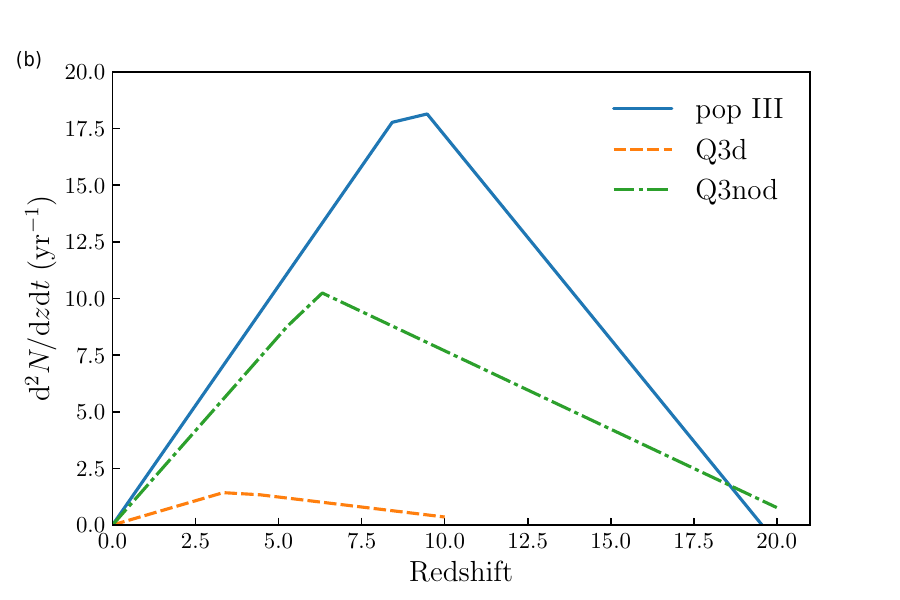}
\caption{{The mass (a) and redshift (b) distributions of MBHBs based on the pop III, Q3d, and Q3nod models used in this work, with $M_z$ defined as $M_z=M(1+z)$.}}\label{fig2}
\end{center}
\end{figure}

In this work, our analysis is based on the inspiral-only GW waveform.
For the non-spinning inspiral of a compact binary system, the GW waveform in frequency-domain is given by
\begin{equation}
\label{eq:hf}
\tilde{h}(f)=\frac{1}{D_{\rm eff}} \sqrt{\frac{5}{24}} \frac{\left(G \mathcal{M}_{\rm chirp}\right)^{5 / 6}}{\pi^{2 / 3} c^{3 / 2}} f^{-7 / 6} \exp(-i\Psi).
\end{equation}
The effective luminosity distance $D_{\rm eff}$ is defined as
\begin{equation}
D_{{\rm eff}}=d_{\rm L} \left[F^{2}_{+}\left(\frac{1+{\rm cos}^2 \iota}{2}\right)^2+F^{2}_{\times} {\rm cos}^2 \iota
\right]^{-1/2},
\end{equation}
where $d_{\rm L}$ is the luminosity distance to the GW source, $\iota$ is the inclination angle between the orbital angular momentum axis of the binary and the line of sight, $G$ is the gravitational constant, $\mathcal{M}_{\rm chirp}=M\eta^{3/5}(1+z)$ is the observed chirp mass, $M=m_1+m_2$ is the total mass of the binary system with component masses $m_1$ and $m_2$, and $\eta=m_1m_2/(m_1+m_2)^2$ is the symmetric mass ratio. The GW phase $\Psi$ is written to the second post-Newtonian order, which is related to the coalescence time $t_{\rm c}$ and the coalescence phase $\psi_{\rm c}$. The detailed form of $\Psi$ can be found in e.g., Ref.~\cite{Ruan:2019tje}. Here, $F_{+,\times}$ are the antenna response functions, which are related to the location of the GW source ($\theta$, $\phi$) and the polarization angle $\psi$. Since the space-based GW observatory could be viewed as two independent interferometers, the $F_{+,\times}$ forms of the second independent interferometer are $F_{+,\times}^{(2)}(t;\theta,\phi,\psi )=F_{+,\times}^{(1)}(t;\theta,\phi-\pi/4,\psi )$. By adopting the low-frequency approximation, the forms of $F_{+,\times}$ for Taiji are from Ref.~\cite{Ruan:2020smc}, for TianQin are from Ref.~\cite{Feng:2019wgq}, and for LISA are from Ref.~\cite{Robson:2018ifk}.
To describe the GW signal in the Fourier space, the observation time $t$ is replaced with $t(f)=t_{\rm c}-\frac{5}{256}\mathcal{M}^{-5/3}_{\rm chirp}(\pi f)^{-8/3}$ \cite{Buonanno:2009zt,Krolak:1995md}. Here we consider Taiji in a heliocentric orbit ahead of the Earth by $20^\circ$, TianQin in a heliocentric orbit around the Earth, and LISA in a heliocentric orbit behind the Earth by $20^\circ$.

\begin{figure*}[!htbp]
\includegraphics[width=\linewidth]{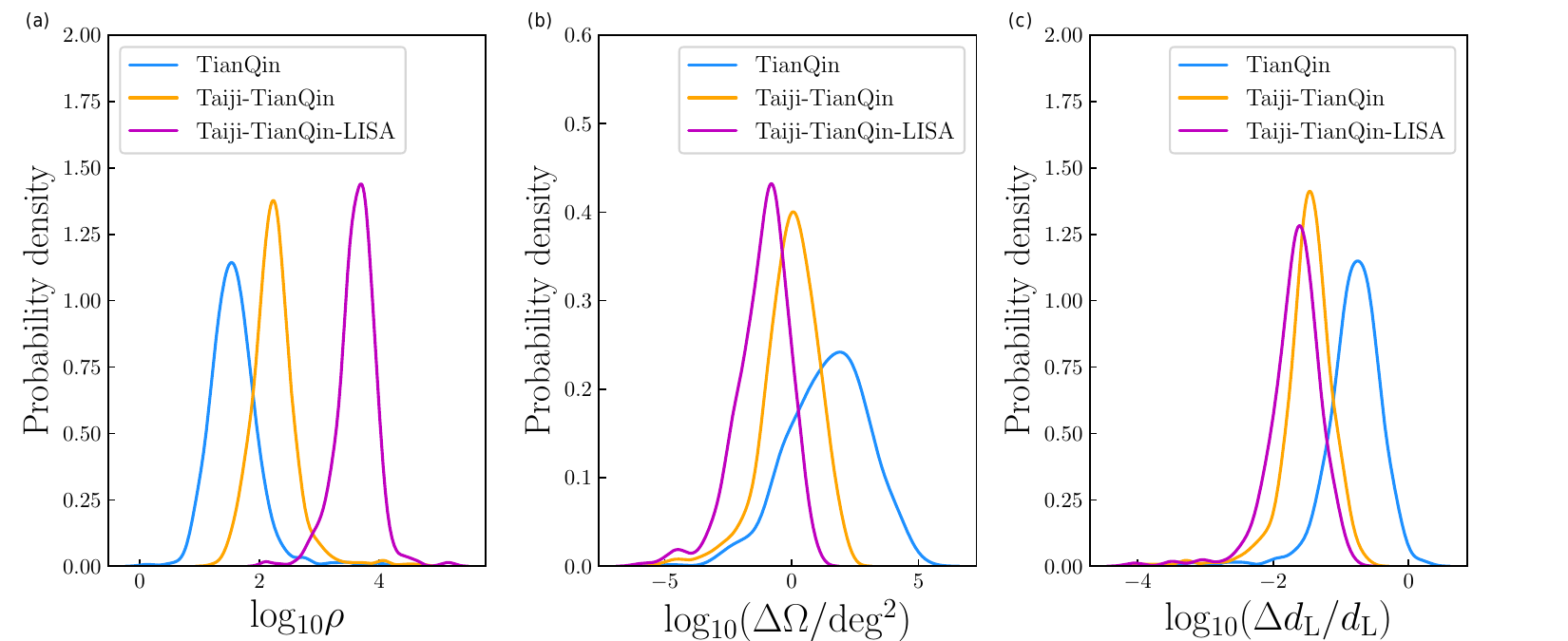}
\centering
\caption{\label{fig3} Distributions of $\rho$, $\Delta\Omega$, and $\Delta d_{\rm L}/d_{\rm L}$ of 500 simulated MBHBs ($z<6$) for TianQin, Taiji-TianQin, and Taiji-TianQin-LISA based on the pop III model. Panels (a)--(c) show the distributions of SNR, sky localization error, and relative uncertainty on luminosity distance, respectively.}
\end{figure*}

\subsection{Parameter estimation of MBHBs}\label{sec:Fisher}
The combined SNR for the detection network of $N$ independent detectors is given by
\begin{equation} \label{SNR}
\rho = ( \tilde{\boldsymbol{h}}\big\vert \tilde{\boldsymbol{h}} ) ^{1/2},
\end{equation}
where $\tilde{\boldsymbol{h}}$ is the frequency domain GW waveform considering the detector network including $N$ independent detectors and can be written as $\tilde{\boldsymbol{h}}=\left[{\tilde{h}_1 },\tilde{h}_2,\cdots,\tilde{h}_k,\cdots,{\tilde{h}_{N}}\right]$. Here $\tilde{h}_k$ is the frequency domain GW waveform of the $k$-th detector.
The inner product is defined as
\begin{align}
({\tilde{\boldsymbol{h}}}\big\vert \tilde{\boldsymbol{h}} )=&\sum_{k=1}^{N} ( \tilde{h}_k\big\vert\tilde{h}_k)=\sum_{k=1}^{N}4\int_{f_{\text {lower}}}^{f_{\mathrm{upper}}}\frac{\tilde{h}_k (f)\tilde{h}^{*}_k(f)}{S_{\text{n},k}(f)} \mathrm{d}f,
\end{align}
where $*$ denotes the complex conjugate, $f_{\rm lower}$ is the lower frequency cutoff, $f_{\rm upper}$ is the upper limit (see Sec.~\ref{subsec:catalog} for more details), and $S_{\text{n},k}(f)$ is the sensitivity curve function of the $k$-th detector when the low-frequency approximation \cite{Robson:2018ifk,Liu:2020eko} is adopted for the response functions $F_{+,\times}$. We adopt the forms of $S_{\text{n}}(f)$ for Taiji from Ref.~\cite{Guo:2018npi}, for TianQin from Ref.~\cite{Wang:2019ryf}, and for LISA from Ref.~\cite{LISACosmologyWorkingGroup:2019mwx}, as shown in Fig.~\ref{fig1}. The SNR threshold is set to be 8 in the simulation.

For a detector network, the Fisher information matrix can be written as
\begin{equation}
\label{fisher}
F_{ij}=\Bigg(\frac{\partial \tilde{\boldsymbol{h}}}{\partial \theta_i}\Bigg\vert \frac{\partial \tilde{\boldsymbol{h}}}{\partial \theta_j}\Bigg)=\sum_{k=1}^{N}\Bigg(\frac{\partial\tilde{h}_{k}}{\partial \theta_i}\Bigg\vert \frac{\partial\tilde{h}_{k}}{\partial \theta_j}\Bigg),
\end{equation}
where $\boldsymbol\theta$ denotes nine parameters ($d_{\rm L}$, $\mathcal{M}_{\rm chirp}$, $\eta$, $\theta$, $\phi$, $\iota$, $t_{\rm c}$, $\psi_{\rm c}$, $\psi$) for a GW event.
The covariance matrix is equal to the inverse of the Fisher matrix, i.e., $\mathrm{Cov}_{ij}=(F^{-1})_{ij}$. Thus the measurement error of GW parameter $\theta_i$ is $\Delta \theta_i =\sqrt{\mathrm{Cov}_{ii}}$. The sky localization error is given by
\begin{equation}
\Delta\Omega=2\pi|\textrm{sin}\theta|\sqrt{(\Delta\theta)^{2}(\Delta\phi)^{2}-(\Delta\theta\Delta\phi)^{2}},
\end{equation}
where $\Delta\theta$, $\Delta\phi$, and $\Delta\theta\Delta\phi$ are given by the covariance matrix.

In addition, the error caused by weak lensing is also considered. We adopt the fitting formula from Refs.~\cite{Hirata:2010ba,Tamanini:2016zlh}
\begin{equation}
\sigma_{d_{\rm L}}^{\rm lens}(z)=d_{\rm L}(z)\times 0.066\bigg[\frac{1-(1+z)^{-0.25}}{0.25}\bigg]^{1.8}.\label{lens}
\end{equation}

{For the dark sirens, the total error of $d_{\rm L}$ is written as
\begin{equation}\label{dls_total_dark}
\sigma_{d_{\rm L}}=\sqrt{(\sigma_{d_{\rm L}}^{\rm inst})^2+(\sigma_{d_{\rm L}}^{\rm lens})^2}.
\end{equation}
Since the redshift of the GW source is unknown when statistically inferring the redshift by using the cross-correlation of MBHBs and galaxy catalogs, we consider the errors of the peculiar velocity of the galaxy and the redshift measurement of the galaxy in the Bayesian analysis, instead of converting them into the luminosity distance errors, as detailedly introduced in Sec.~\ref{subsec:dark}. The redshift error caused by the peculiar velocity of the galaxy is given by
\begin{equation}
  \sigma_{z}^{\rm pv}(z)=(1+z)\frac{\sqrt{\langle v^2\rangle}}{c},
\end{equation}
where $\sqrt{\langle v^{2}\rangle}$ is the peculiar velocity of the galaxy. In this work, we set $\sqrt{\langle v^{2}\rangle}=500$ km $\cdot$ s$^{-1}$, in agreement with the average value observed in galaxy catalogs \cite{He:2019dhl}.
In addition, we make the assumption that the redshift of the galaxy is measured spectroscopically with neglected error at $z\leq 1$. While for the galaxy at $z>1$, the redshift is measured photometrically with error. Such treatment is also adopted in Ref.~\cite{Zhu:2021aat}.
The redshift error caused by the redshift measurement $\sigma_{z}^{\rm meas}$ predicted for the Vera Rubin Observatory in the redshift range of $0<z<4$ must be smaller than $0.05(1+z)$, with a goal of $0.02(1+z)$ \cite{LSSTScience:2009jmu}. Therefore, we consider the medium redshift error, i.e., $\sigma_{z}^{\rm meas}\approx 0.03(1+z)$ to make the following analysis. Hence, the total redshift error of the galaxy is given by
\begin{equation}
    \sigma_{z}(z)=\left\{\begin{array}{cc}
\begin{aligned}
    &(1+z)\frac{\sqrt{\langle v^2\rangle}}{c},&&z\leq 1,\\
    &(1+z)\frac{\sqrt{\langle v^2\rangle}}{c}+0.03(1+z), &&z>1.
    \end{aligned}
    \label{eq:PhiL}
\end{array}\right.
\end{equation}}

For the bright sirens, we additionally consider the luminosity distance errors caused by the peculiar velocity $\sigma_{d_{\rm L}}^{\rm pv}$ and the redshift measurements $\sigma^{z}_{d_{\rm L}}$.
The error caused by the peculiar velocity of the GW source is given by \cite{Kocsis:2005vv}
\begin{equation}
\sigma_{d_{\rm L}}^{\rm pv}(z)=d_{\rm L}(z)\times \bigg[1+ \frac{c(1+z)^2}{H(z)d_{\rm L}(z)}\bigg]\frac{\sqrt{\langle v^2\rangle}}{c}.
\end{equation}
Following Refs.~\cite{Tamanini:2016zlh,Speri:2020hwc,Wang:2021srv}, we propagate the redshift errors to the distance errors by assuming our fiducial cosmology,
\begin{equation}
\sigma^{z}_{d_{\rm L}}=\frac{\partial d_{\rm L} }{\partial z}\Delta z.
\end{equation}
Hence, the total error of $d_{\rm L}$ for the bright siren is written as
\begin{equation}
\sigma_{d_{\rm L}}=\sqrt{(\sigma_{d_{\rm L}}^{\rm inst})^2+(\sigma_{d_{\rm L}}^{\rm lens})^2+(\sigma_{d_{\rm L}}^{\rm pv})^2+(\sigma^{z}_{d_{\rm L}})^2}.\label{dls_total_bright}
\end{equation}
Due to the perfect sky localizations of the bright sirens, we can conduct the follow-up EM observations. Hence, we can convert the errors caused by the peculiar velocity and the redshift measurement into the luminosity distance error.
Meanwhile, we make the assumption that redshift measurements at $z<2$ are determined spectroscopically with negligible errors, which is also adopted in Refs.~\cite{Speri:2020hwc,Wang:2019tto,Zhao:2019gyk,Wang:2021srv}. However, for the GW events with $z>2$ associated with photometric measurements \cite{Speri:2020hwc,Dahlen:2013fea}, the redshift errors are estimated as $\Delta z\approx 0.03(1+z)$ \cite{Ilbert:2013bf,Dahlen:2013fea}. {Note that for the bright siren method, the redshift information can be measured from the follow-up EM observations, instead of obtaining from the galaxy survey catalog. Therefore, in our analysis, the spectroscopic redshift measurement range for the bright sirens is larger than that for the dark sirens.}

\subsection{Galaxy localization}\label{subsec:Galaxy}

We first obtain the localization errors and luminosity distance errors of MBHBs. Meanwhile, we simulate the galaxy catalogs uniformly in a co-moving volume with a number density of $0.02$ Mpc$^{-3}$ \cite{Barausse:2012fy} in the range of $z\in [0, 3]$ (the adopted galaxy number density is in the middle of the observational error bars; see Fig.~1 of Ref.~\cite{Barausse:2012fy}). By combining the flat priors for $\Omega_{\rm m}\in [0.1, 0.5]$ and $H_0\in [60, 80]\ {\rm km\ s^{-1}\ Mpc^{-1}}$, and the luminosity distance errors, we obtain the lower and upper limits of redshift ($z_{\rm min}$, $z_{\rm max}$), assuming the $\Lambda$CDM model. We then derive the $2\times 2$ covariance matrix $\mathrm{Cov}[\theta, \phi]$ from the whole covariance matrix and use the new covariance matrix to calculate $\chi^2$, which describes the angular deviation from an arbitrary galaxy to the GW source and is given by
\begin{equation}
\chi^{2}=\boldsymbol{\xi}\left(\operatorname{Cov}^{-1}\right) \boldsymbol{\xi}^{\top}=\sum_{i, j} \xi_{i}\left(\operatorname{Cov}^{-1}\right)_{i j} \xi_{j},\label{chi2}
\end{equation}
with $\boldsymbol{\xi}=(\theta-\bar{\theta},\phi-\bar{\phi})$. Here ($\bar{\theta}$, $\bar{\phi}$) represent the angular location of the GW source and ($\theta$, $\phi$) represent the angular location of an arbitrary galaxy. We adopt the boundary of the GW source's angular localization with $\chi^2=9.21$, corresponding to the 99\% confidence level. Therefore, if the galaxies satisfy $\chi^2\leq 9.21$ and $z\in [z_{\rm min}, z_{\rm max}]$, they can be considered as the potential host galaxies of the GW source.

\section{Simulations}\label{sec:simulation}

\subsection{Simulations of GW catalogs}\label{subsec:catalog}

In this subsection, we detailedly introduce the simulations of MBHB catalogs. Following Refs.~\cite{Klein:2015hvg,Tamanini:2016zlh,Wang:2021srv}, we consider both the ``light-seed'' and ``heavy-seed'' scenarios for the seeding models of MBHB \cite{Barausse:2012fy,Madau:2001sc,Volonteri:2007ax}.
First, we would briefly introduce the seeding models of MBHB considered in this work.

(i) Model pop III: a light-seed scenario that assumes the MBH seed is the remnants of the first generation or population III stars, with a mass of around $100~M_{\odot}$ \cite{Madau:2001sc,Volonteri:2002vz}.

(ii) Model Q3d: a heavy-seed scenario that assumes the MBH seed is from the collapse of protogalactic disks \cite{Klein:2015hvg}, with a mass of $\sim10^5~M_{\odot}$, and also considers the time delay between the merger of MBHs and merger of galaxies.

(iii) Model Q3nod: same as Q3d, but not consider the time delay.

According to the merger rates of MBHBs predicted in Ref.~\cite{Klein:2015hvg}, for the simulation of bright sirens, we simulated 877, 41, and 610 GW catalogs based on the pop III, Q3d, and Q3nod models. {In Fig.~\ref{fig2}, we show the fitting distributions of $z$ and $M_z$ considered in this work.}
For the simulation of dark sirens, following Ref.~\cite{Zhu:2021aat}, we make a reasonable assumption that it is difficult to obtain complete galaxy catalogs at $z>3$, so we only select the GW events at $z < 3$ for the three models of MBHB as the dark siren events and simulate the galaxy catalogs at $z < 3$. 

In this work, we consider the following observation strategies for space-based GW detectors:
\begin{itemize}
\item{Taiji}: a space-based GW observatory consists of three satellites forming a triangle with the arm length of $3\times 10^6$ km, and the mission time is 5 years \cite{Guo:2018npi,Wu:2018clg,Hu:2017mde}.
\item{TianQin}: a space-based GW observatory consists of two constellations, each consisting of an equilateral triangle of three satellites with the arm length of $\sqrt{3}\times 10^5$ km \cite{Liu:2020eko,Wang:2019ryf,Luo:2015ght,Luo:2020bls,Milyukov:2020kyg,TianQin:2020hid}, and the mission time is 5 years.
\item{LISA}: a space-based GW observatory consists of three satellites forming a triangle with the arm length of $2.5\times 10^6$ km  \cite{Robson:2019,LISA:2017pwj,LISACosmologyWorkingGroup:2022jok}, and the mission time is 5 years\footnote{In fact, the lifetime of LISA is commonly considered to be 4 years. However, LISA is also prepared to extend to 10 years \cite{LISA:2017pwj}. Furthermore, recent work has shown that extending the mission to 6 years would greatly benefit scientific investigations \cite{Seoane:2021kkk}.}.
\item{Taiji+TianQin}: Taiji and TianQin observe together, with 5 years of overlap in operation time.
\item{Taiji+TianQin+LISA}: Taiji, TianQin, and LISA observe together, with 5 years of overlap in operation time\footnote{We assume that the duty cycle during the operation time is 100\%, which means that the space-based GW detectors are continuously observing in the operation time. However, recent work shows that the duty cycle of LISA is only about 0.75 \cite{Seoane:2021kkk}. This indicates that it may take about 6.7 years of observation time to achieve the results in our analysis. Nonetheless, since Taiji, TianQin, and LISA all have the potential to extend their observation time, we consider this assumption to be reasonable.}.
\end{itemize}

The sky location ($\theta$, $\phi$), the binary inclination angle $\iota$, the polarization angle $\psi$, the coalescence time $t_{\rm c}$, and the coalescence phase $\psi_{\rm c}$ are randomly sampled in the ranges of ${\rm cos}\,\theta\in [-1,1]$, $\phi\in [0,2\pi]$, ${\rm cos}\,\iota\in [-1,1]$, $\psi\in [0,2\pi]$, $t_{\rm c}\in [0,5]$ years, and $\psi_{\rm c}\in [0,2\pi]$, respectively.
For Taiji, TianQin, and LISA, we adopt the frequency in the range of $[10^{-4}, 1]$ Hz. In the calculations of Eqs.~(\ref{SNR}) and (\ref{fisher}), $f_{\rm lower}$ is chosen as $f_{\rm lower}=\max(f_{\rm obs},10^{-4})$ Hz and $f_{\rm upper}$ is chosen as $f_{\rm upper}=\min(f_{\rm ISCO},1)$ Hz. Here $f_{\rm obs}=(t_{\rm c}/5)^{-3/8}\mathcal{M}_{\rm chirp}^{-5/8}/8\pi$ is the observation frequency at the coalescence time $t_{\rm c}$ and $f_{\rm ISCO}$ is the innermost stable circular orbit (ISCO) frequency $f_{\rm upper}=c^{3}/6\sqrt{6}\pi GM_{\rm obs}$ with $M_{\rm obs}=(m_{1}+m_{2})(1+z)$ \cite{Feng:2019wgq}.

\subsection{Bright sirens: observations of EM counterparts}\label{subsec:EM}

\begin{figure}[htbp]
\begin{center}
\includegraphics[width=7.2cm,height=5.4cm]{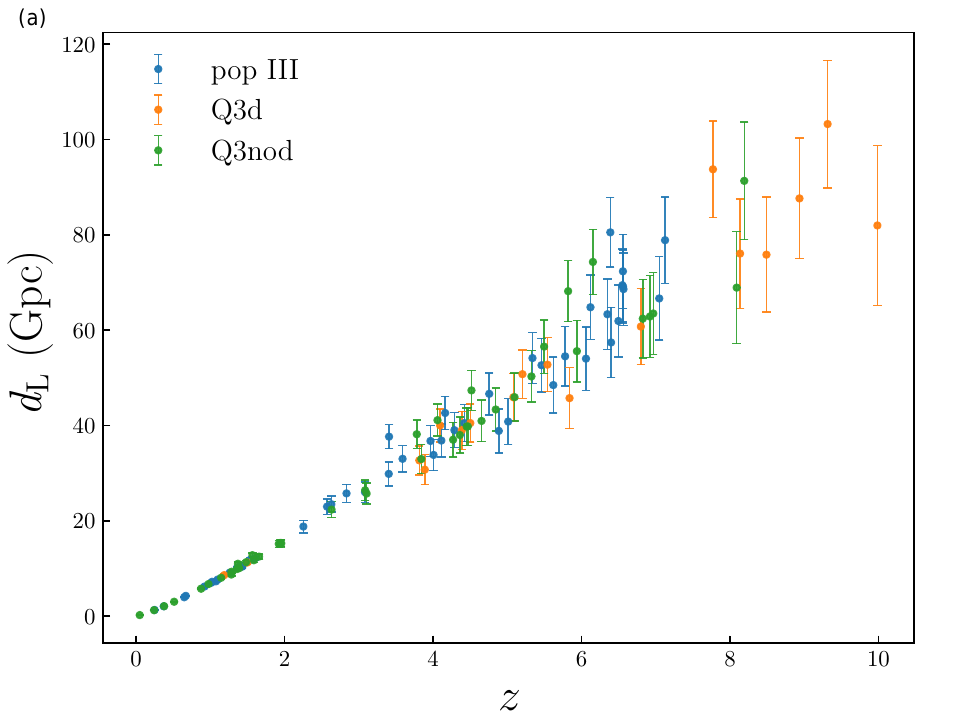}\\
\includegraphics[width=7.2cm,height=5.4cm]{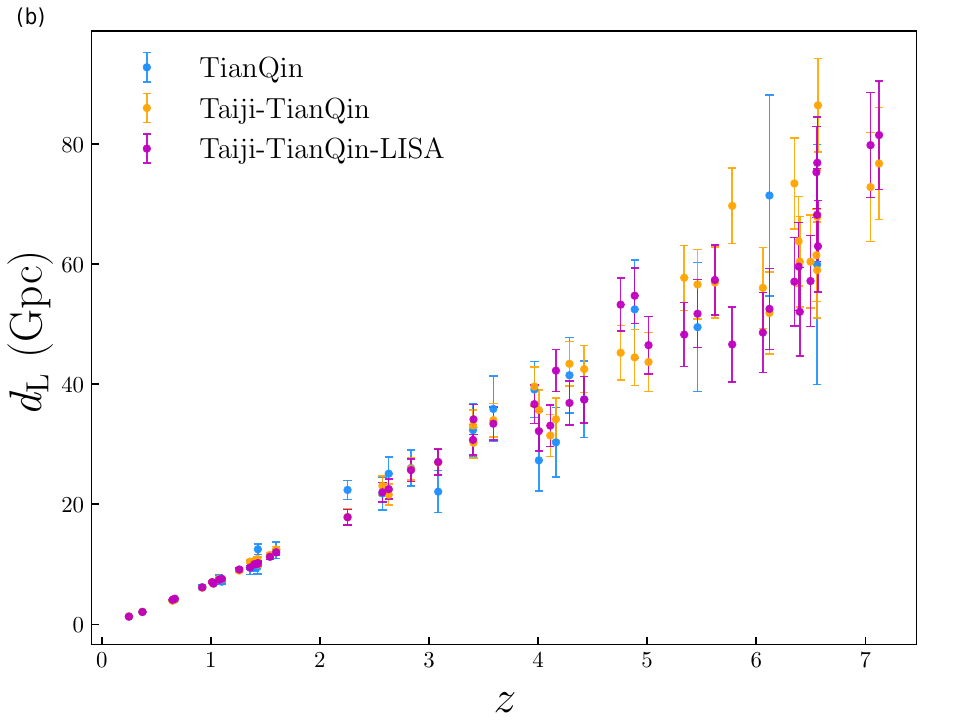}
\end{center}
\caption{The simulated bright siren data. Panel (a): the simulated bright sirens from the 5-year observation of the Taiji-TianQin-LISA network based on the pop III, Q3d, and Q3nod models. Panel (b): the simulated bright sirens from the 5-year observation of TianQin, the Taiji-TianQin network, and the Taiji-TianQin-LISA network based on the pop III model.}
\label{fig4}
\end{figure}

\begin{table*}[htbp]
\setlength\tabcolsep{2.8pt}
\renewcommand{\arraystretch}{1.5}
\caption{\label{tab1} The estimated detection rates of GW-EM events from TianQin, the Taiji-TianQin network, and the Taiji-TianQin-LISA network within 5-year operation time based on the pop III, Q3d, and Q3nod models.}
\centering
\resizebox{\textwidth}{!}
{
\begin{tabular}{lccccccccc}
\\
\hline\hline  &\multicolumn{2}{c}{pop III}&&\multicolumn{2}{c}{Q3d}&& \multicolumn{2}{c}{Q3nod}\\
 \cline{2-3}\cline{5-6}\cline{8-9}
 Detection strategy&Detection percentage&Number&  &Detection percentage&Number&   &Detection percentage&Number\\
\hline

TianQin

                  &$3.2\%$
                  &$28$&
                  &$24.3\%$
                  &$8$&
                  &$2.1\%$
                  &$13$\\
Taiji-TianQin

                 &$5.6\%$
                 &$49$&
                 &$39\%$
                 &$16$&
                 &$5.9\%$
                 &$36$\\
Taiji-TianQin-LISA
                &$5.6\%$
                &$49$&
                &$44\%$
                &$18$&
                &$7\%$
                &$43$\\
\hline
\hline
\end{tabular}}
\centering
\end{table*}

Several works suggest that the mergers of MBHBs could emit EM radiations in the radio/optical band \cite{Palenzuela:2010nf,OShaughnessy:2011nwl,Moesta:2011bn,Kaplan:2011mz,Shi:2011us,Blandford:1977ds,Meier:2000wk,Dotti:2011um} (see Refs.~\cite{Gold:2014dta,Farris:2014zjo} for models which can emit EM radiations continuously in the inspiral phase). Redshift measurements are crucial for using MBHBs as standard sirens. To detect EM radiations from MBHBs, radio and optical telescopes must be aimed at the GW sources in advance. A conservative method is to select GW events with $\rho\geq8$ and $\Delta\Omega\leq10\ \text{deg}^2$ (corresponding to the field of view of LSST) during the inspiral-only phases of MBHBs as potential candidates for EM counterpart detections\footnote{We assume that real-time data analysis of the GW source position during the inspiral phase is feasible in the era of space-based GW observatories, thanks to the potential of deep learning algorithms \cite{George:2017pmj,George:2016hay,Gebhard:2019ldz}.}. If the GW source can be located to better than $10\ \text{deg}^2$, it could provide early warning for radio/optical telescopes, such as SKA and ELT, thus greatly enhancing the detection of EM counterparts (see Appendix \ref{Appendix:warning time} for more details about the warning time).

In Fig.~\ref{fig3}, we present the distributions of $\rho$, $\Delta\Omega$, and $\Delta d_{\rm L}/d_{\rm L}$ for TianQin, Taiji-TianQin, and Taiji-TianQin-LISA based on the pop III model. These distributions are calculated by 500 random simulations in the redshift range of $[0,6]$. We can see that the SNRs of Taiji-TianQin-LISA are the highest, followed by Taiji-TianQin and TianQin. For the sky localization errors, compared to the single TianQin observatory, the Taiji-TianQin network could greatly reduce sky localization errors. With the addition of the LISA observatory, the localization ability could be improved to a certain extent. For the measurement precisions of $d_{\rm L}$, the above results still hold.

Previous work has shown that the EM counterparts detected by LSST are also detectable for SKA+ELT \cite{Tamanini:2016zlh}. Therefore, we consider the strategy of SKA+ELT to observe EM counterparts in the radio and optical bands. We assume that the characterization of EM emission consists of radio flares and jets, and optical luminosity flares. In this case, the EM counterparts may first be detected by SKA in the radio band, and the host galaxies are then identified through localization. Then, the redshifts are determined spectroscopically or photometrically by the optical telescope ELT. The total luminosity in the radio band is given by $L_{\rm radio}=L_{\rm flare}+L_{\rm jet}$ (detailed forms of $L_{\rm flare}$ and $L_{\rm jet}$ can refer to Refs.~\cite{Palenzuela:2010nf,Moesta:2011bn,OShaughnessy:2011nwl,Kaplan:2011mz,Shi:2011us,Tamanini:2016zlh}). The luminosity that can be detected by SKA needs to satisfy the following relationship
\begin{equation}
\left(\frac{L_{\rm radio}}{\rm erg/s}\right) \left(\frac{d_{\rm L}}{\rm cm}\right)^{-2} \geq 4\pi\, 10^{-18}\, \left(\frac{F_{\nu,{\rm min}}^{\rm SKA}}{\rm\mu Jy}\right) \left(\frac{\nu_{\rm SKA}}{\rm GHz}\right). \label{L_radio}
\end{equation}
Following Ref.~\cite{OShaughnessy:2011nwl}, we assume that the bulk of the emission takes place in the SKA band, and here we have $\nu_{\rm SKA}\simeq1.4~{\rm GHz}$ and $F^{\rm SKA}_{\nu,{\rm min}}\simeq1~\mu{\rm Jy}$\footnote{The current goal of SKA phase 1 mid-frequency is a flux limit of $F^{\rm SKA}_{\nu,{\rm min}}\simeq2~\mu{\rm Jy}$ on a $0.5~\rm deg^2$ field of view assuming the integration time of 10 minutes (Table~1 of Ref.~\cite{SKA-design}). Considering the high performance of full SKA, we therefore assume that $F^{\rm SKA}_{\nu,{\rm min}}\simeq1~\mu{\rm Jy}$ on approximately $10~\rm deg^2$, in the same integration time \cite{Tamanini:2016zlh}.}. Eq.~(\ref{L_radio}) assumes that radio waves are emitted isotropically. Actually, the situation is not so straightforward, as the synchrotron emission resulting from particle acceleration within the jet should be collimated along the jet direction. The jet's opening angle can cause a decrease in luminosity. However, if the jet is aligned towards us, we would receive a signal with a greater flux, which can lead to the detection of a dimmer source that would be otherwise undetectable. Since the two factors have opposite effects on the number of the GW-EM events and may counteract each other. Therefore, we adopt a simple approximation to treat the radio flux isotropically, which is also adopted in Refs.~\cite{Tamanini:2016zlh,Wang:2021srv}.

The redshift information cannot be obtained from the identification of radio emission. Therefore, follow-up observations using ELT are required to search for EM emissions in the optical band and to determine the redshift. If the EM radiation in the optical band produced by the merger of MBHB satisfies the following expression, the redshift can be measured by ELT
\begin{align}
m_{\rm gal}=82.5-\frac{5}{2}{\rm log}_{10}\left(\frac{L_k}{3.02}\frac{{\rm s}}{{\rm erg}}\right)+5{\rm log}_{10}\left(\frac{d_{\rm L}}{{\rm pc}}\right)\leq m_{{\rm ELT}},
\end{align}
where $m_{\rm gal}$ is the apparent magnitude which is calculated by the host galaxy luminosity and $L_k$ is the galaxy luminosity in the $K$-band. Here we follow Refs.~\cite{Tamanini:2016zlh,Wang:2021srv} and assume a fiducial mass-to-light ratio $M/L_k=0.03$ in the simulation (in fact, $M/L_k$ for young stellar populations at moderate redshift is in the range of $0.01-0.05$ \cite{Bruzual:2003tq}). $m_{{\rm ELT}}=31.3$ is the detection threshold of ELT, which is the photometric limiting magnitude of ELT corresponding to the $J$-band and $H$-band. In principle, the threshold should be set to 30.2 corresponding to the $K$-band \cite{MICADOTeam:2010oam}, because the apparent magnitude is calculated by the host galaxy's luminosity in $K$-band. Since MICADO (Multi-AO Imaging Camera for Deep Observations) on ELT will cover the range of 1000--2400 nm ($J$-band to $K$-band), we simply set the highest limiting magnitude as the threshold. Although this approach may slightly overestimate the detection threshold, it does not significantly affect the number of simulated bright sirens. In Refs.~\cite{Speri:2020hwc,Wang:2021srv}, the authors adopt a simple method of taking into account the redshift error for all the GW events at $z>2$, because the spectroscopic redshift at $z>2$ is usually unavailable \cite{Dahlen:2013fea,Speri:2020hwc}. In fact, as shown in Fig.~\ref{fig4}, most simulated GW-EM events are at $z>2$, i.e., the redshift errors are considered for most data points. On the other hand, the errors from lensing for the GW events at $z>2$ are dominant. Hence, such a treatment has only a minor impact on the cosmological parameter estimations.

The estimated detection rates of GW-EM events from TianQin, Taiji-TianQin, and Taiji-TianQin-LISA are shown in Table~\ref{tab1}. We can see that the number of detected GW-EM events from the Taiji-TianQin network almost double compared to the single TianQin observatory. For the heavy-seed models of MBHB, Q3d, and Q3nod, the addition of LISA could improve the number of detected GW-EM events. While for the light-seed model, pop III, the addition of LISA has no improvement for the number of detected GW-EM events. As shown in the middle panel of Fig.~\ref{fig3}, the prime cause is that in the pop III model, the sky localization errors of the Taiji-TianQin network are good enough, and thus the addition of LISA could only improve the localization ability to some extent. For the Taiji-TianQin-LISA network, the detection rates of the pop III, Q3d, and Q3nod models are 5.6\%, 44\%, and 7\%, respectively.

In Fig.~\ref{fig4}, we show the simulated bright siren data. In panel (a), we show the bright sirens from the Taiji-TianQin-LISA network based on the pop III, Q3d, and Q3nod models. We can see that for the Q3d model, the redshifts of the bright sirens could reach $z\sim 10$. In panel (b), we can see that the Taiji-TianQin-LISA network could detect the most number of bright sirens, followed by the Taiji-TianQin network and TianQin, also clearly shown in Table~\ref{tab1}. Moreover, the detector network could detect higher-redshift bright sirens than the single TianQin observatory.

We use the simulated bright siren data to perform cosmological analysis. We adopt the Markov Chain Monte Carlo (MCMC) method \cite{Lewis:2002ah} to maximize the likelihood $\mathcal{L}$ and infer the posterior probability distributions of cosmological parameters.

For the discussions about bright sirens, we adopt the pop III model of MBHB as the representative of the MBHB model. This is because (i) the pop III model typically offers intermediate constraints among the three models and (ii) the pop III model is less affected by the inspiral-merger-ringdown (IMR) waveform (while considering the IMR waveform has a significant correction for the heavy-seed model) \cite{Tamanini:2016zlh}, i.e., the number of GW-EM events based on the light-seed pop III model is less affected by the IMR waveform. Therefore, we use the pop III model to make discussions in the following.

We emphasize that our analysis is conservative because we adopt the GW waveform in the inspiral phase. As mentioned above, it is commonly believed that EM signals are emitted in the mergers of MBHBs. Hence, the warning time of EM counterparts is important for the application of the standard siren method (see Appendix \ref{Appendix:warning time} for more details). In addition, in order to obtain robust constraint results of cosmological parameters, we randomly simulate 40 sets of bright siren data and adopt the data which give the medium constraint results as the final used bright siren data.

\subsection{Dark sirens: Bayesian method}\label{subsec:dark}

\begin{figure}[!htbp]
\includegraphics[width=0.5\textwidth]{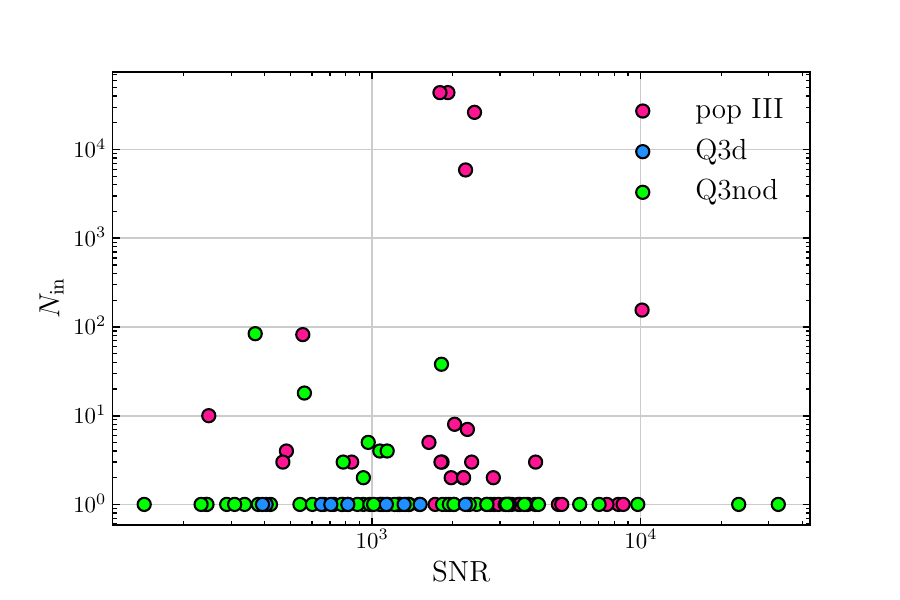}
\centering
\caption{Correlation between SNR of the dark siren from the Taiji-TianQin-LISA network and the number of the potential host galaxies $N_{\rm in}$ based on the pop III, Q3d, and Q3nod models.}\label{fig5}
\end{figure}

For the GW events that the redshifts cannot be obtained by identifying EM counterparts, we need to use the cross-correlation of MBHBs and galaxy catalogs to obtain redshift information. Throughout this work, we apply a redshift cutoff for events beyond $z=3$ and assume that the galaxy catalog at $z<3$ is complete, which is also adopted in e.g., Ref.~\cite{Zhu:2021aat}.
{In fact, it is challenging to obtain the complete galaxy catalogs at high redshift. The Dark Energy Survey \cite{DES:2017myt} and Sloan Digital Sky Survey (SDSS) \cite{SDSS-III:2015hof,SDSS:2017bih,SDSS-IV:2019txh} can reliably map the galaxies at $z\leq 1.2$. Nonetheless, future EM survey projects are expected to obtain higher-redshift complete galaxy catalogs. Meanwhile, the idea that quasars can host MBHBs is discussed in e.g., Refs.~\cite{Sobacchi:2016yez,Connor:2019gun,Penil:2020uqg}. The SDSS survey can map the quasar catalog at $z\simeq 4$ \cite{SDSS:2017bih,SDSS-IV:2019txh}, which makes it possible to detect MBHBs at higher redshift. Moreover, once MBHBs at higher redshift are detected, intensive observations will be expected to be triggered to map higher-redshift galaxy catalogs within the sky localization error region.
In addition, one can also explore the potential of obtaining the redshift information using the Lyman-$\alpha$ forest effect \cite{Hess:2018qit,Ravoux:2020bpg,Qin:2021gkn,Porqueres:2019dpn}. The above facts show that it is possible to obtain the complete galaxy catalogs at $z<3$.}
In this subsection, we briefly introduce the method of using the dark siren method to infer the cosmological parameters.
Using the method introduced in Sec.~\ref{subsec:Galaxy}, we calculate the number of the potential host galaxies $N_{\rm in}$ and show the SNR--$N_{\rm in}$ plot in Fig.~\ref{fig5}.

We adopt the Bayesian method to infer the cosmological parameters and the posterior probability distribution of the cosmological parameters $\vec{\Omega}$ is
\begin{equation}
  p(\vec{\Omega}|\vec{X}_{\rm GW},\vec{D}_{\rm GW})\propto p(\vec{\Omega})\mathcal{L}(\vec{X}_{\rm GW}|\vec{D}_{\rm GW},\vec{\Omega}),
\end{equation}
where $\vec{X}_{\rm GW}$ represent the GW data, $\vec{D}_{\rm GW}$ represent that the GW events are detected,
and $p(\vec{\Omega})$ are the prior distributions of the cosmological parameters.
Assuming that the observations of GW events are independent of each other, the likelihood of the GW data can be expressed as
\begin{equation}
    \mathcal{L}(\vec{X}_{\rm GW}|\vec{D}_{\rm GW},\vec{\Omega})=\prod_{i=1}^{N_{\rm GW}}\mathcal{L}(X_{{\rm GW},i}|D_{{\rm GW},i},\vec{\Omega}),
\end{equation}
where $N_{\rm GW}$ is the number of GW events. The likelihood function $\mathcal{L}(X_{{\rm GW}}|D_{{\rm GW}},\vec{\Omega})$ is obtained by marginalizing the redshifts of the GW events,
\begin{equation}
    \mathcal{L}(X_{\rm GW}|D_{\rm GW},\vec{\Omega})=\frac{\int p(X_{\rm GW}|d_{\rm L}(z,\vec{\Omega}))p_{z}(z){\rm d}z }{\int p(D_{\rm GW}|d_{\rm L}(z,\vec{\Omega}))p_{z}(z){\rm d}z },
\end{equation}
where $p(X_{\rm GW}|d_{\rm L}(z,\vec{\Omega}))$ is the posterior distribution of GW event's $d_{\rm L}$ obtained from the GW observation
and is given by
\begin{equation}
\label{con:GW likelihood}
    \begin{aligned}
    p\left( X_{\rm GW}|d_{\rm L}(z,\vec{\Omega})\right)=\frac{1}{\sqrt{2\pi}\sigma _{d_{\rm L}}}\exp\left[-\frac{(\hat{d}_{\rm L}-d_{\rm L}(z,\vec{\Omega}))^2}{2\sigma _{d_{\rm L}}^2}\right ],
    \end{aligned}
\end{equation}
where $\hat{d}_{\rm L}$ is the observed luminosity distance of the GW source and the luminosity distance error $\sigma_{d_{\rm L}}$ can be obtained from Eq.~(\ref{dls_total_dark}).
{$p_{z}(z)$ is the redshift distribution of the GW source, which is obtained by combining the EM and the GW observations and can be expressed as
\begin{equation}
    \begin{aligned}
        p_{z}(z)\propto\frac{1}{N_{\rm in}}\sum_{n=1}^{N_{\rm in}}W_{n}G(\hat{z}_{n},z,\sigma_{z,n})p_0(z),
    \end{aligned}
\end{equation}
where $G(\hat{z}_{n},z,\sigma_{z,n})$ is the Gaussian function describing the redshift distribution of the $n$-th potential host galaxy, and $\hat{z}_n$ and $\sigma_{z,n}$ are the expected value and the 1-$\sigma$ error of the redshift, respectively, which is discussed in Sec.~\ref{sec:Fisher}.}
$W_{n}$ is the angular position weight of the $n$-th potential host galaxy
\begin{equation}
    W_n\propto \exp\left(-\frac{1}{2}\chi^2_n\right).
\end{equation}
Here $\chi^2_n$ is obtained from Eq.~(\ref{chi2}). {$p_0(z)$ is the prior redshift distribution of galaxies, and we assume it to be a uniform distribution within a comoving volume, shown as follows,
\begin{equation}
    \begin{aligned}
        p_0(z)\propto\frac{{\rm d}V_{\rm c}}{{\rm d}z}\propto\frac{d_{\rm c}^2(z)}{H(z)},
    \end{aligned}
\end{equation}
where $V_{\rm c}$ is the comoving volume and $d_{\rm c}(z)$ is the comoving distance with respect to redshift $z$.}

Following Ref.~\cite{Gray:2019ksv}, the likelihood function's denominator describes the GW observation's selection effects. $p(D_{\rm GW}\vert d_{\rm L}(z,\vec{\Omega}))$ is the probability of detecting the GW event at $d_{\rm L}(z,\vec{\Omega})$ and is obtained by marginalizing the other GW source parameters.
\begin{align}
    p(D_{\rm GW}\vert d_{\rm L}&(z,\vec{\Omega}))\nonumber=\int p(D_{\rm GW}\vert d_{\rm L}(z,\vec{\Omega}),X_{\rm GW}){\rm d}X_{\rm GW}\nonumber\\&=\frac{1}{N_{\rm sample}}\sum\limits_{i=1}^{N_{\rm sample}}p(D_{{\rm GW},i}\vert d_{\rm L}(z,\vec{\Omega}),X_{{\rm GW},i}),
\end{align}
where $N_{\rm sample}$ is set to 50000 and $X_{{\rm GW},i}$ represents the $i$-th GW event whose luminosity distance is $d_{\rm L}(z,\vec{\Omega})$ and the other GW source parameters are randomly chosen according to their prior distributions. $p(D_{{\rm GW},i}|d_{\rm L}(z,\vec{\Omega}),X_{{\rm GW},i})$ is the probability of detecting the $i$-th GW event, expressed as
\begin{equation}
    \begin{aligned}
        p(D_{\mathrm{GW}, i} \big\vert d_{\mathrm{L}}(z, \vec{\Omega})), X_{\mathrm{GW}, i})=\left\{\begin{array}{ll}
1, & \text { if } \rho_{i}>\rho_{\mathrm{th}}, \\
0, & \text { otherwise. }
\end{array}\right.
    \end{aligned}
\end{equation}

\begin{table*}[!htbp]
\setlength\tabcolsep{3pt}
\renewcommand{\arraystretch}{1.5}
\caption{\label{tab2} {The absolute errors ($1\sigma$) and the relative errors of the cosmological parameters in the $\Lambda$CDM, $w$CDM, and $w_0w_a$CDM models using the CMB, Taiji-TianQin-LISA (pop III), and CMB+Taiji-TianQin-LISA (pop III) data. Here, $H_0$ is in units of ${\rm km\ s^{-1}\ Mpc^{-1}}$.}}
\centering
\resizebox{\textwidth}{!}
{
\begin{tabular}{ccccccccccccccc}
\hline \multirow{2}{*}{Error} &\multicolumn{3}{c}{$\Lambda$CDM}&& \multicolumn{3}{c}{$w$CDM}&& \multicolumn{3}{c}{$w_{0}w_{a}$CDM}\\
 \cline{2-4}\cline{6-8}\cline{10-12}
  &CMB & Taiji-TianQin-LISA&CMB+Taiji-TianQin-LISA&&CMB  &Taiji-TianQin-LISA  &CMB+Taiji-TianQin-LISA&&CMB &Taiji-TianQin-LISA  &CMB+Taiji-TianQin-LISA\\
\hline
$\sigma(\Omega_{\rm m})$
                   & $0.009$
		           &$0.019$
                   & $0.005$&
                    & $0.064$
                   & $0.027$
                    &$0.007$&
                    &$0.060$
                    &$0.056$
                    &$0.017$\\

$\sigma(H_0)$
                   & $0.61$
		           &$0.62$
                   & $0.38$&
                   & $7.30$
                   & $1.40$
                    &$0.71$&
                    &$6.10$
                    &$2.00$
                    &$1.70$\\

$\sigma(w)$
                   & $-$
                   & $-$
                   & $-$&
                   & $0.250$
                   & $0.200$
                    &$0.036$&
                    &$-$
                    &$-$
                    &$-$\\

$\sigma(w_0)$
                   & $-$
                   & $-$
                   & $-$&
		           &$-$
                   & $-$
                    &$-$&
                    &$0.610$
                    &$0.330$
                    &$0.230$\\

$\sigma(w_a)$
                   & $-$
                    &$-$
                   & $-$&
		           &$-$
                    &$-$
                    &$-$&
                    &$-$
                    &$1.70$
                    &$0.63$\\
\hline

$\varepsilon(\Omega_{\rm m})$
                   & $2.7\%$
		           &$5.8\%$
                   & $1.7\%$&
                   & $20.5\%$
                   & $8.5\%$
                    &$2.0\%$&
                    &$18.6\%$
                    &$15.5\%$
                    &$5.3\%$\\

$\varepsilon(H_0)$
                    & $0.9\%$
		           &$0.9\%$
                   & $0.6\%$&
                   & $10.6\%$
                   & $2.1\%$
                    &$1.1\%$&
                    &$9.0\%$
                    &$3.0\%$
                    &$2.5\%$\\

$\varepsilon(w)$
                   & $-$
                   & $-$
                   & $-$&
                   & $23.8\%$
                   & $18.9\%$
                    &$3.6\%$&
                    &$-$
                    &$-$
                    &$-$\\

$\varepsilon(w_0)$
                   & $-$
                   & $-$
                   & $-$&
		           &$-$
                   & $-$
                    &$-$&
                    &$98.3\%$
                    &$36.7\%$
                    &$23.4\%$\\

\hline
\end{tabular}
}
\centering
\end{table*}

\section{Cosmological parameter estimation}\label{sec:Result}

\begin{figure*}[!htbp]
\begin{center}
\includegraphics[width=0.4\textwidth]{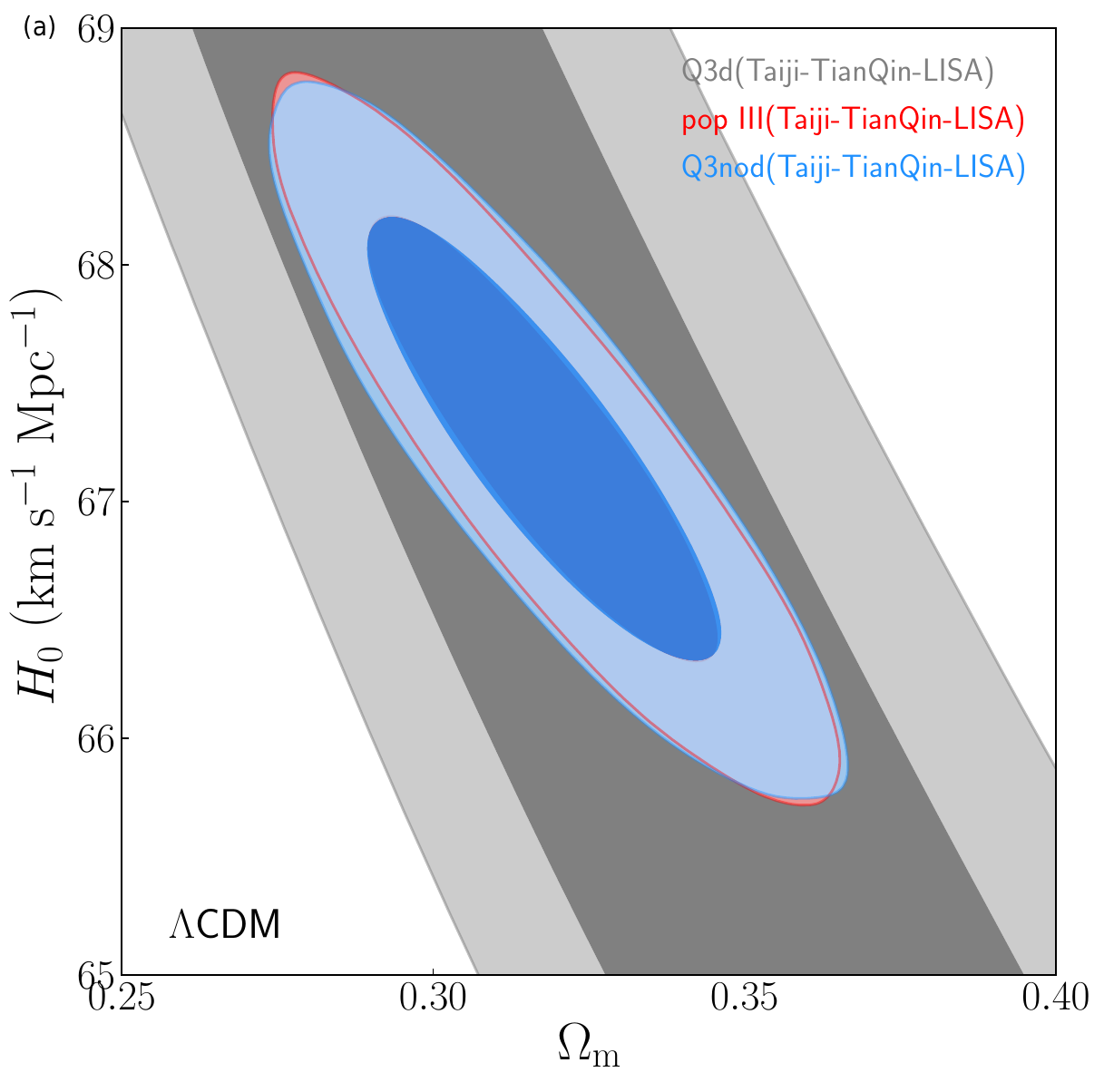}
\includegraphics[width=0.4\textwidth]{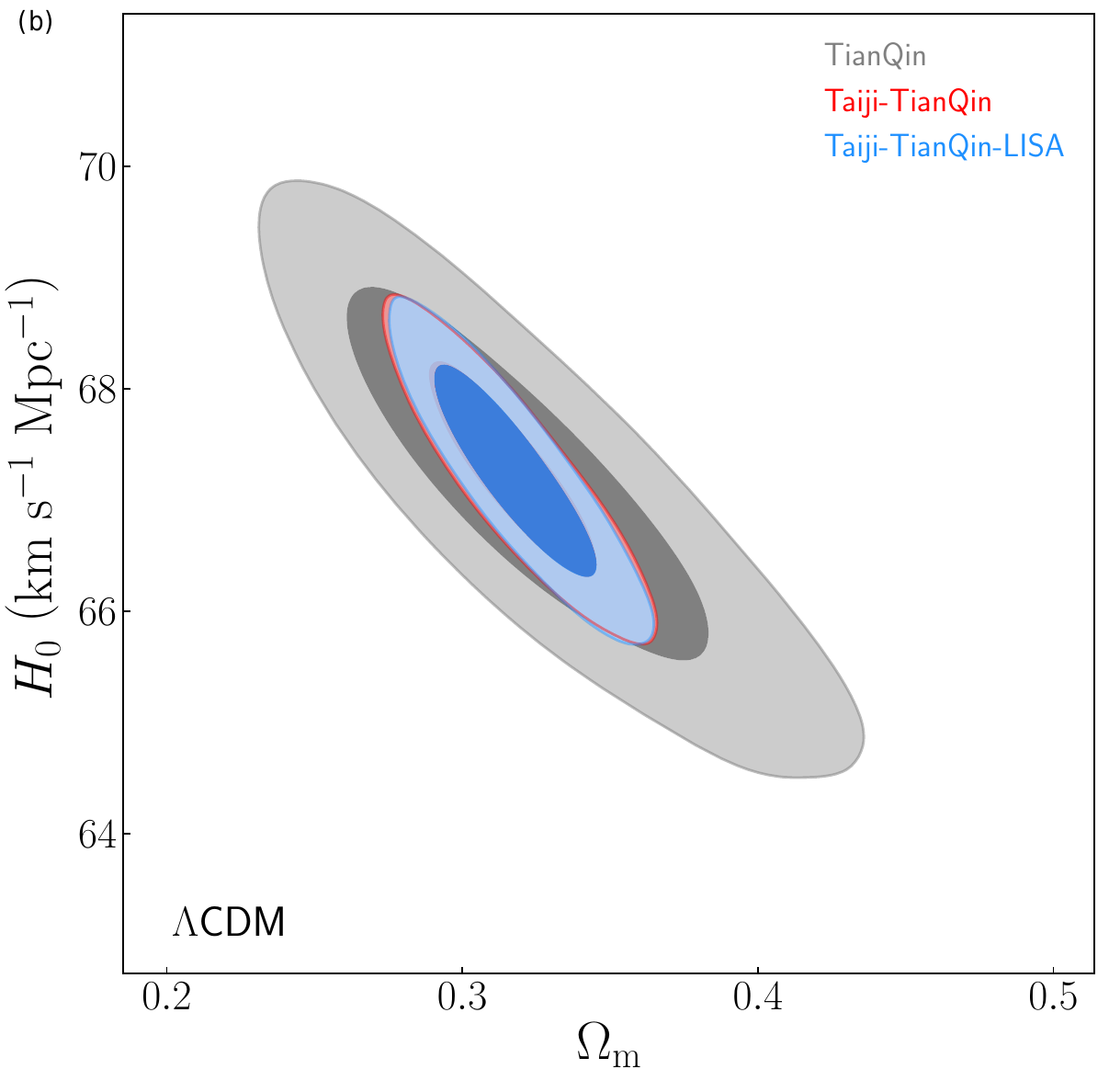}
\caption{Constraints on the $\Lambda$CDM model. Panel (a): Two-dimensional marginalized contours ($68.3\%$ and $95.4\%$ confidence level) in the $\Omega_{\rm m}$--$H_0$ plane using the simulated bright siren data from the Taiji-TianQin-LISA network based on the pop III, Q3d, and Q3nod models. Panel (b): Two-dimensional marginalized contours ($68.3\%$ and $95.4\%$ confidence level) in the $\Omega_{\rm m}$--$H_0$ plane using the simulated bright siren data from TianQin, the Taiji-TianQin network, and the Taiji-TianQin-LISA network based on the pop III model.}\label{fig6}
\end{center}
\end{figure*}

\begin{figure*}[!htb]
\begin{center}
\includegraphics[width=0.4\linewidth,angle=0]{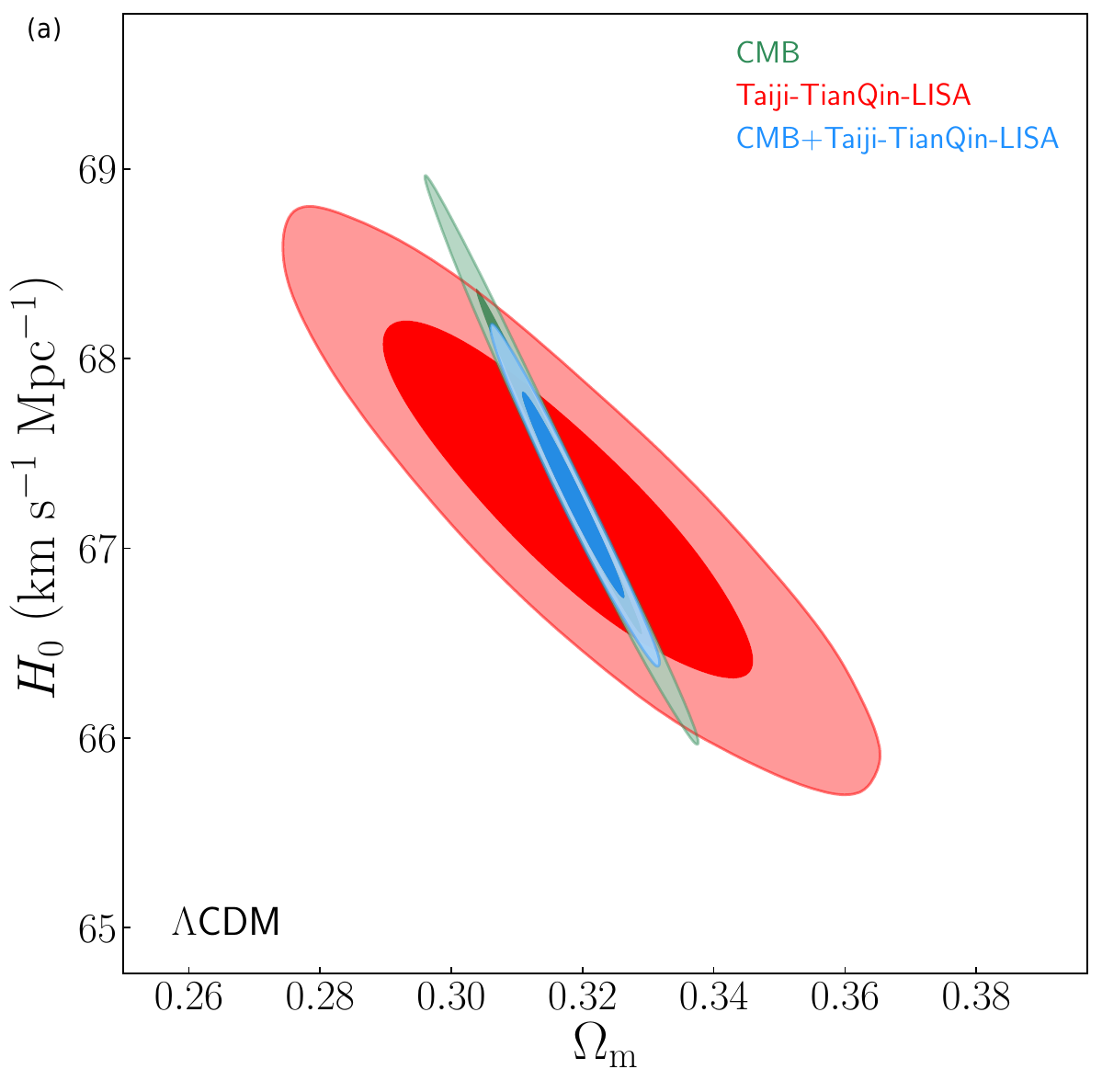} \hspace*{.0cm}
\includegraphics[width=0.4\linewidth,angle=0]{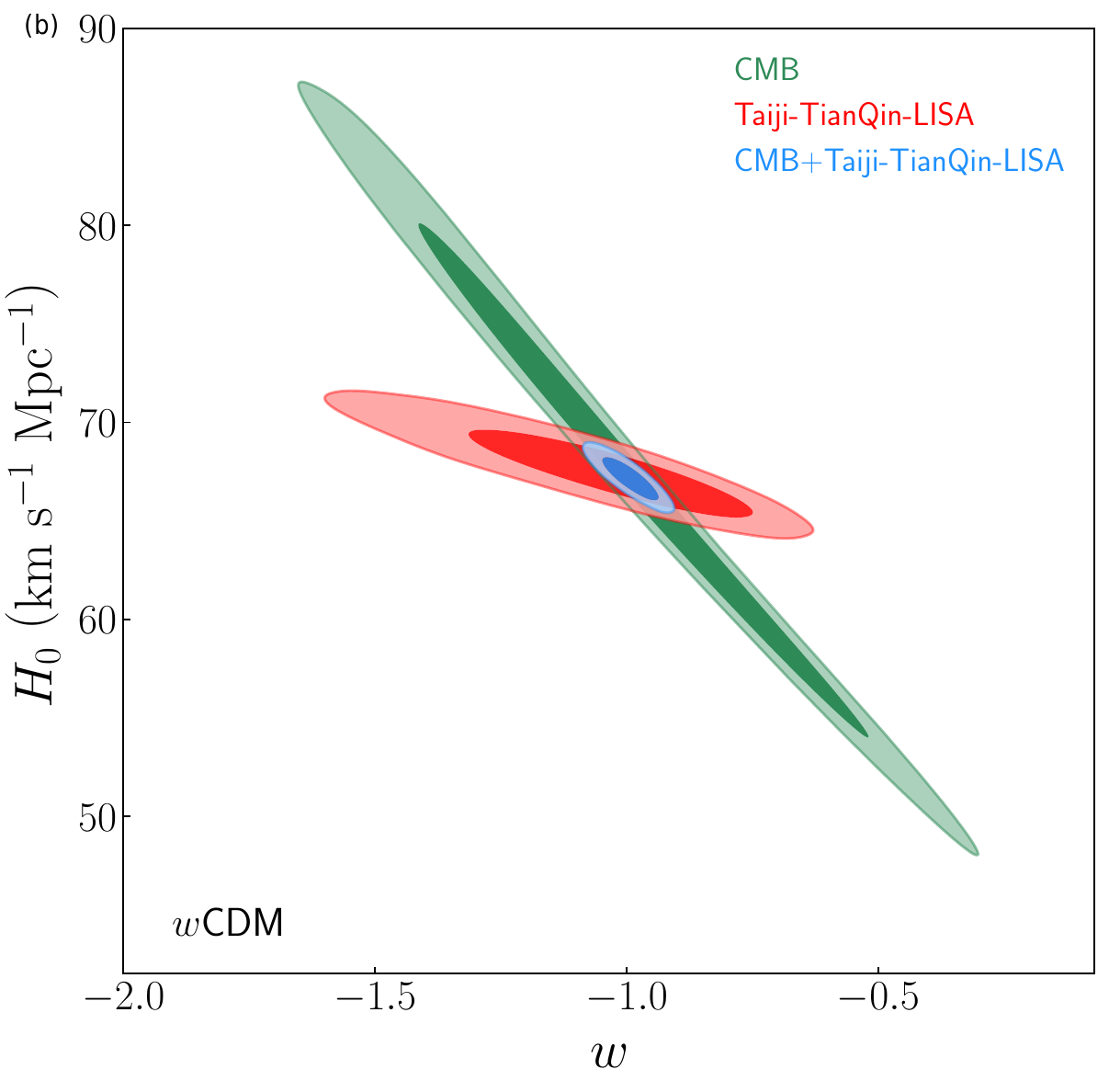}\\
\includegraphics[width=0.4\linewidth,angle=0]{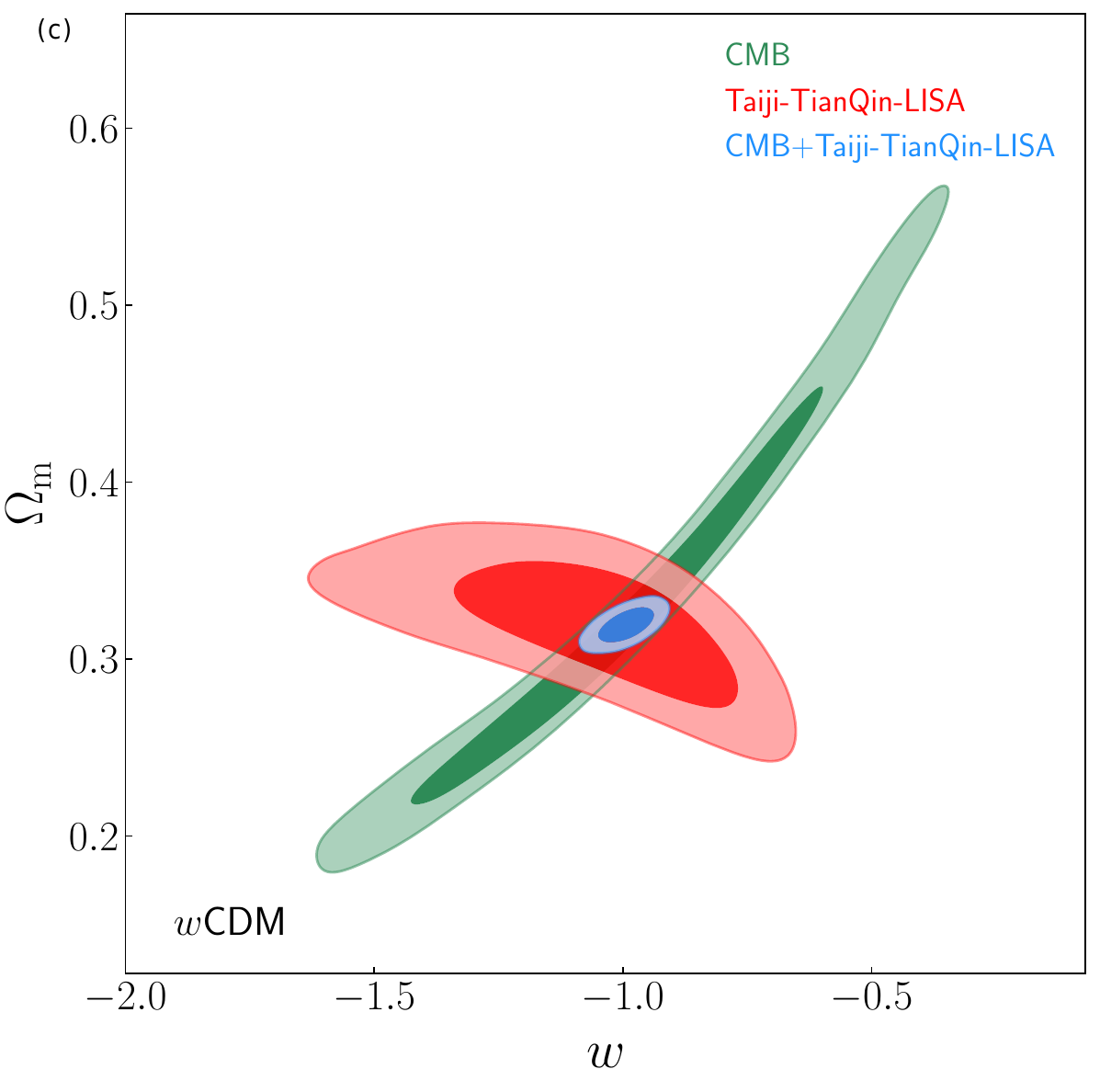}
\includegraphics[width=0.4\linewidth,angle=0]{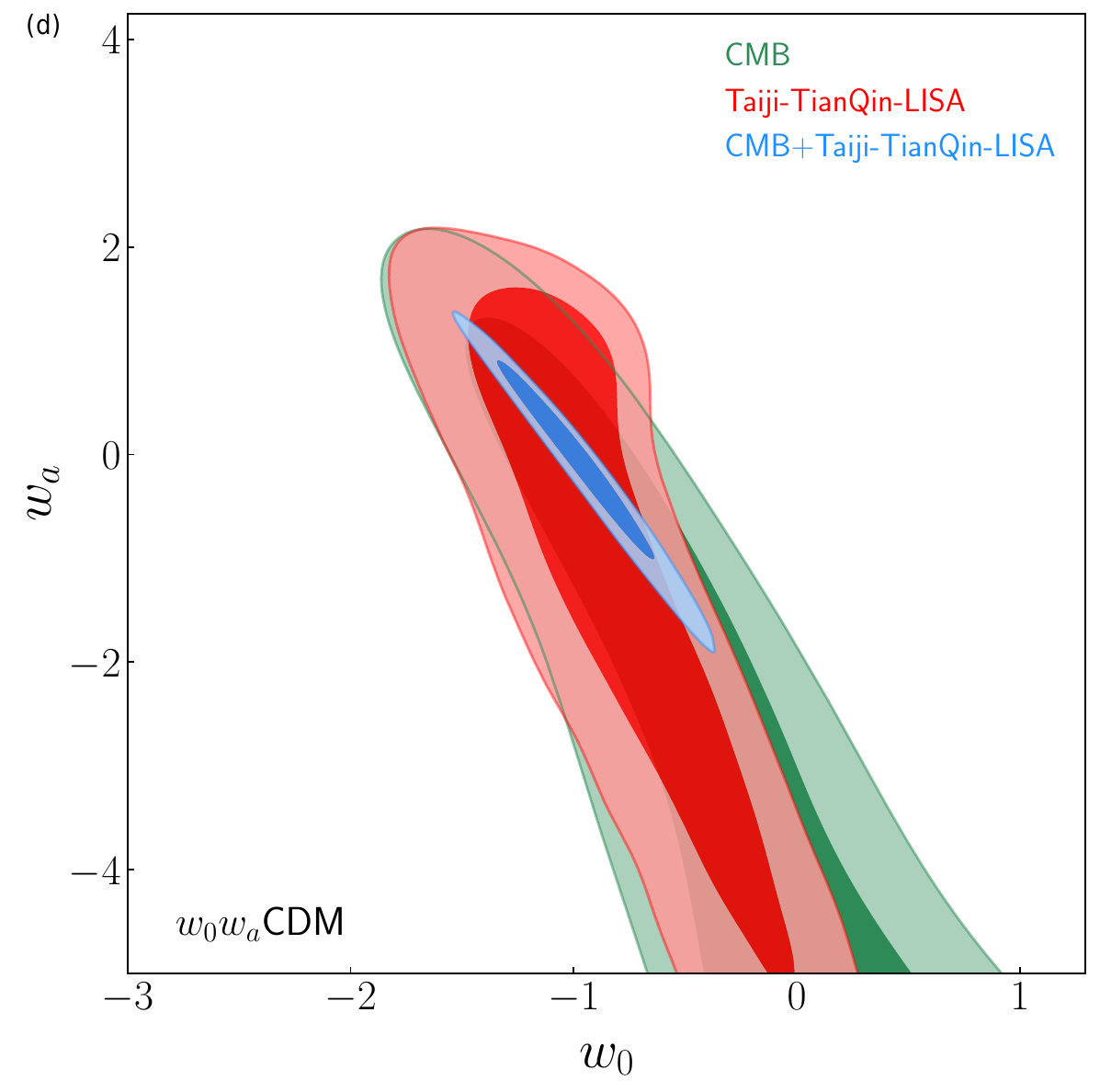}
\end{center}
\caption{Constraints on the $\Lambda$CDM, $w$CDM, and $w_0w_a$CDM models using the CMB, Taiji-TianQin-LISA, and CMB+Taiji-TianQin-LISA data. Panel (a): Two-dimensional marginalized contours (68.3\% and 95.4\% confidence level) in the $\Omega_{\rm m}$--$H_{0}$ plane for the $\Lambda$CDM model using the CMB, Taiji-TianQin-LISA, and CMB+Taiji-TianQin-LISA data. Panel (b): Similar to panel (a), but in the $w$--$H_0$ plane for the $w$CDM model. Panel (c): Similar to panel (a), but in the $w$--$\Omega_{m}$ plane for the $w$CDM model. Panel (d): Similar to panel (a), but in the $w_0$--$w_a$ plane for the $w_0w_a$CDM model. The Taiji-TianQin-LISA data are simulated based on the pop III model.}\label{fig7}
\end{figure*}

\begin{figure}[!htb]
\begin{center}
\includegraphics[width=\linewidth,angle=0]{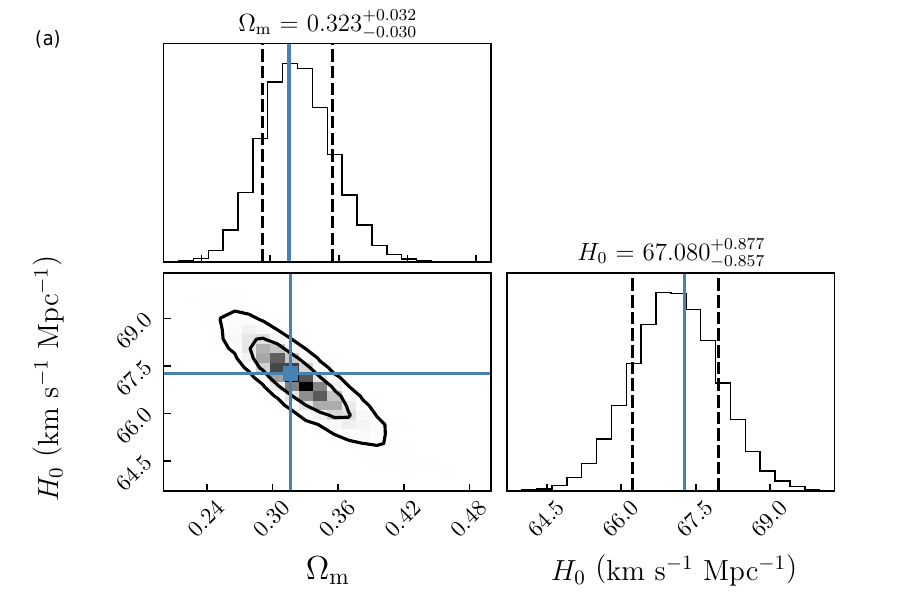} \hspace*{.0cm}
\includegraphics[width=\linewidth,angle=0]{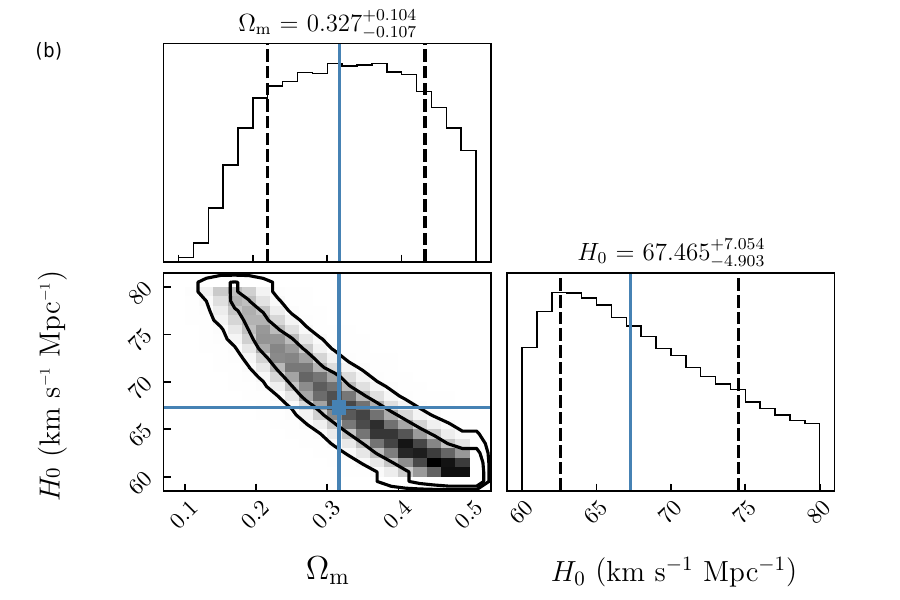}\\
\includegraphics[width=\linewidth,angle=0]{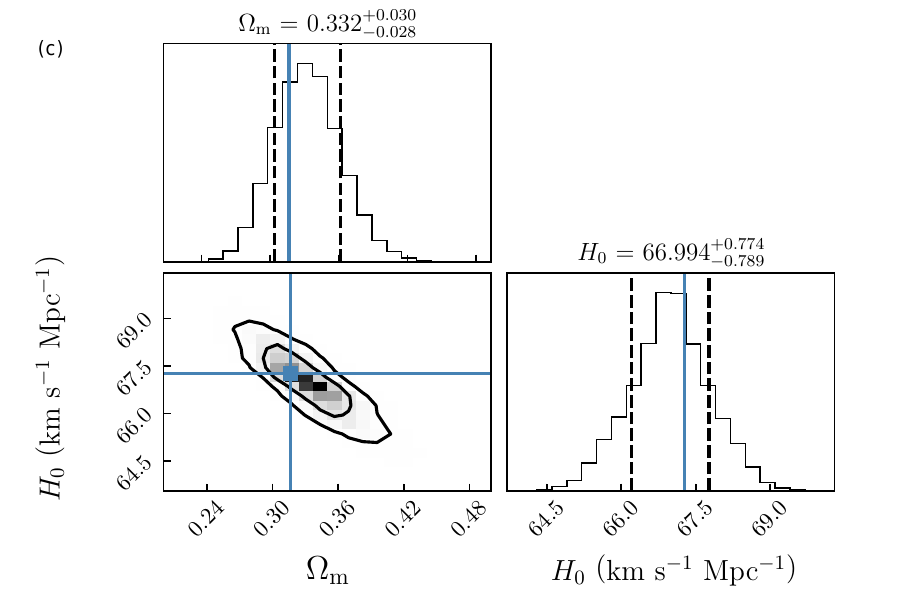}
\end{center}
\caption{Constraints on the $\Lambda$CDM model using the mock dark siren data from the Taiji-TianQin-LISA network based on the pop III (a), Q3d (b), and Q3nod (c) models. Here the galaxy catalogs are uniformly simulated in a co-moving volume with a number density of $0.02$ Mpc$^{-3}$ \cite{Barausse:2012fy} in the range of $z\in [0, 3]$.}\label{fig8}
\end{figure}

In this section, we shall report the constraint results of the cosmological parameters. We consider the optimistic (bright sirens) and conservative (dark sirens) scenarios to make the cosmological analysis. For the optimistic case, we use the simulated bright siren data from the Taiji-TianQin-LISA network to constrain the $\Lambda$CDM, $w$CDM, and $w_0w_a$CDM models by performing the Markov-chain Monte Carlo analysis \cite{Lewis:2002ah} {based on the $\tt emcee$ Python module \cite{Foreman-Mackey:2012any}}. In order to show the ability of the Taiji-TianQin-LISA network to break cosmological parameter degeneracies, we also show the constraint results of CMB and CMB+Taiji-TianQin-LISA.
For the CMB data, we employ the ``Planck distance priors'' from the Planck 2018 observation \cite{Chen:2018dbv,Planck:2018vyg}. The $1\sigma$ and $2\sigma$ posterior distribution contours for the cosmological parameters of interest are shown in Figs.~\ref{fig6} and \ref{fig7} and the $1\sigma$ errors for the marginalized parameter constraints are summarized in Table \ref{tab2}. We also calculate the constraint results of TianQin and the Taiji-TianQin network for comparison. Other constraint results are shown in Appendix \ref{Appendix:cosmological parameter}. We use $\sigma(\xi)$ and $\varepsilon(\xi)$ to represent the absolute and relative errors of parameter $\xi$, with $\varepsilon(\xi)$ defined as $\varepsilon(\xi)=\sigma(\xi)/\xi$. Note that we adopt the pop III model as the representative of the MBHB model to show the ability of the Taiji-TianQin-LISA network to break cosmological parameter degeneracies, as discussed in Sec.~\ref{subsec:catalog}.
For the conservative case, we use the dark sirens from the Taiji-TianQin-LISA network to constrain the $\Lambda$CDM model. The constraint results are shown in Fig.~\ref{fig8} and summarized in Table~\ref{tab3}.

\subsection{Bright sirens}

In Fig.~\ref{fig6}(a), we show the constraint results in the $\Omega_{\rm m}$--$H_{0}$ plane for the $\Lambda$CDM model using the simulated standard sirens from the Taiji-TianQin-LISA network based on the three population models of MBHB. As can be seen, Q3d gives the worst constraint results due to the least number of standard sirens. The detection number of pop III is more than that of Q3nod. However, the pop III and Q3nod models give similar constraint results. The prime cause is that Q3nod has more lower-redshift data points ($z<2$), so Q3nod (43 data points) gives comparable constraint results with those of pop III (49 data points). In Fig.~\ref{fig6}(b), in order to clearly show the ability of the Taiji-TianQin-LISA network to constrain cosmological parameters, we show the constraints using the simulated standard siren data from TianQin, the Taiji-TianQin network, and the Taiji-TianQin-LISA network based on pop III. We could see that Taiji-TianQin-LISA gives the best constraints, followed by Taiji-TianQin and TianQin. Compared with TianQin, the Taiji-TianQin network could significantly improve the constraint precisions of cosmological parameters. With the addition of LISA to the Taiji-TianQin network, the constraint precisions of cosmological parameters are improved slightly. In the scenario of pop III, the Taiji-TianQin network gives $\sigma(\Omega_{\rm m})=0.019$ and $\sigma(H_0)=0.63$ km s$^{-1}$ Mpc$^{-1}$, which are 53.7\% [(0.041-0.019)/0.041] and 42.7\% [(1.10-0.63)/1.10] better than those of TianQin. In fact, the Taiji-TianQin network could constrain the Hubble constant to a precision of 0.9\%. Meanwhile, the Taiji-TianQin-LISA network gives almost the same constraints on $\Omega_{\rm m}$ and $H_0$.

In Fig.~\ref{fig7}, we show the constraints on the $\Lambda$CDM, $w$CDM, and $w_0w_a$CDM models using the CMB, Taiji-TianQin-LISA (pop III), and the CMB+Taiji-TianQin-LISA (pop III) data. We first focus on the $\Lambda$CDM model. From Fig.~\ref{fig7}(a), we can see that the contours of CMB and Taiji-TianQin-LISA show different orientations and thus the combination of them could break the parameter degeneracies. The combination of CMB and Taiji-TianQin-LISA gives $\sigma(\Omega_{\rm m})=0.005$ and $\sigma(H_0)=0.38$ km s$^{-1}$ Mpc$^{-1}$, which are 44.4\% and 37.7\% better than those of CMB. In Figs.~\ref{fig7}(b) and \ref{fig7}(c), we show the cases in the $w$CDM model. We can see that the contours of CMB and Taiji-TianQin-LISA also show different orientations. In particular, the parameter degeneracy orientations in the $w$--$\Omega_{\rm m}$ are almost orthogonal. Hence, the combination of them could effectively break the cosmological parameter degeneracies and greatly improve the constraint precisions of cosmological parameters. The combination of CMB and Taiji-TianQin-LISA gives $\sigma(\Omega_{\rm m})=0.007$, $\sigma(H_0)=0.71$ km s$^{-1}$ Mpc$^{-1}$, and $\sigma(w)=0.036$, which are 89.1\%, 90.3\%, and 85.6\% better than those of CMB. Moreover, the constraint precision of $w$ is 3.6\%, which is close to the latest constraint results by the CMB+SN data \cite{Brout:2022vxf}. In Fig.~\ref{fig7}(d), we show the constraint results in the $w_0w_a$CDM model. Also, the combination of CMB and Taiji-TianQin-LISA could break cosmological parameter degeneracies. The combination of CMB and Taiji-TianQin-LISA gives $\sigma(w_0)=0.23$ which is 62.3\% better than the constraint by the CMB data.

\subsection{Dark sirens}

\begin{table}
\setlength\tabcolsep{4pt}
\renewcommand{\arraystretch}{1.5}
\caption{Constraint results of the dark sirens from the Taiji-TianQin-LISA network based on the pop III, Q3d, and Q3nod models. The first column represents the MBHB model, the second column represents the number of the simulated dark siren data at $z<3$ used in this work, the third column represents the percentages of GW events with $N_{\text {in}}=1$, the fourth column represents the relative errors of $\Omega_{\rm m}$, and the fifth column represents the relative errors of $H_0$.}
\begin{tabular}{ccccccc}
\hline \hline
MBHB model & Total number &$N_{\text {in }}=1$ &  $\varepsilon(\Omega_{\rm m})$& $\varepsilon(H_0)$ \\
\hline pop III & 47 & {$55.3 \%$}  &  {$9.6\%$} & {$1.29\%$} \\
\hline Q3d & 9 & {$100 \%$}  &  {$32.3\%$} & {$8.86\%$} \\
\hline Q3nod & 52 & {$89.4 \%$}  &  {$8.7\%$} & {$1.17\%$} \\
\hline \hline \label{tab3}
\end{tabular}
\end{table}

\begin{table}
\setlength\tabcolsep{10pt}
\renewcommand{\arraystretch}{1.5}
\caption{Constraint results of the dark sirens from the Taiji-TianQin network, the Taiji-LISA network, and the TianQin-LISA network based on the pop III model.}
\begin{tabular}{ccccccc}
\hline \hline
Detection strategy & & $\varepsilon(\Omega_{\rm m})$& &$\varepsilon(H_0)$ \\
\hline Taiji-TianQin &  &$14.05\%$ & &$2.70\%$ \\
\hline Taiji-LISA & &  $10.77\%$ & &$1.82\%$ \\
\hline TianQin-LISA & & $13.79\%$ && $2.73\%$ \\
\hline \hline \label{tab4}
\end{tabular}
\end{table}

In Fig.~\ref{fig8}(a), we show the constraints in the $\Omega_{\rm m}$--$H_0$ plane for the pop III model. We can see that the Taiji-TianQin-LISA network gives $\sigma(\Omega_{\rm m})=0.031$ and $\sigma(H_0)=0.867\ {\rm km\ s^{-1}\ Mpc^{-1}}$, which are all worse than the constraint results by the simulated bright sirens from the Taiji-TianQin-LISA network. However, the constraint precision of $H_0$ is 1.29\%, close to the standard of precision cosmology. In Fig.~\ref{fig8}(b), we show the case for the Q3d model. The Taiji-TianQin-LISA network gives poor constraints $\sigma(\Omega_{\rm m})=0.106$ and $\sigma(H_0)=5.979\ {\rm km\ s^{-1}\ Mpc^{-1}}$ due to the low merger rate of MBHB. In Fig.~\ref{fig8}(c), we show the case for the Q3nod model. In this case, the Taiji-TianQin-LISA network gives similar constraint results to those of the pop III model. The constraint precision of $H_0$ is 1.17\%, also almost meeting the standard of precision cosmology.

{For comparison, in Table~\ref{tab4}, we also show the constraint results of the Taiji-TianQin network, the Taiji-LISA network, and the TianQin-LISA network (the detailed contours are shown in Appendix~\ref{Appendix:cosmological parameter}). We can see that Taiji-LISA gives the best constraint results, while Taiji-TianQin and TianQin-LISA give similar constraint results. The constraint precision of the Hubble constant using Taiji-LISA is 1.82\%, which is worse than the result given in Ref.~\cite{Wang:2020dkc}. The prime cause is that our results additionally consider the redshift errors of the galaxy caused by peculiar velocity and the redshift measurement, leading to looser constraints. While the constraint precision of the Hubble constant using TianQin-LISA is consistent with that given in Ref.~\cite{Zhu:2021aat}.
The addition of the LISA observatory to the Taiji-TianQin network can improve the constraint on the Hubble constant by 51.9\%.
Meanwhile, the addition of the TianQin observatory to the Taiji-LISA network can improve the constraint on the Hubble constant by 28.9\%.
Although the two-detector network and the three-detector network show similar constraint results in constraining cosmological parameters in the bright siren scenario, the three-detector network can improve the localization ability compared to the two-detector network. Therefore, in the dark siren scenario, the three-detector network can improve the constraint on the Hubble constant to some extent.}

{Meanwhile, we can see that the measurement precision of $H_0$ using dark sirens of the space-based detector network is much better than 19\% given by the current dark siren measurement. The prime cause is that the localization ability of the space-based detector network is much better than that of the LIGO-Virgo-KAGRA network, with $\Delta \Omega$ of the LIGO-Virgo-KAGRA network mainly distributing in the order of $\mathcal{O}(100)~\mathrm{deg}^2$ \cite{LIGOScientific:2020ibl,LIGOScientific:2021djp}.
In addition, the completeness of the current $\tt GLADE+$ galaxy catalog is about 20\% up to 800 Mpc \cite{LIGOScientific:2021aug}. In this work, we consider the next-generation galaxy survey projects and assume that the galaxy catalog at $z<3$ is complete.}

{Here we highlight three points in the dark siren analysis. First, we assume that all the MBHB mergers are dark. Actually, the dark siren events with good localization $N_{\rm in}=1$ have the potential to be upgraded to bright sirens. Second, we do not account for the effect of galaxy clustering. Taking the galaxy clustering effect into consideration can lead to a more concentrated redshift distribution, which could help reduce the measurement error of $H_0$ \cite{MacLeod:2007jd,Mukherjee:2020hyn}. Third, we refrain from utilizing any galaxy properties other than redshift in our analysis, because the current understanding of the relationship between MBHs and galaxies is still uncertain. Furthermore, unlike stellar binary black holes, the population of MBHB is small. Due to the good localization of the space-based detector network, the EM telescopes could conduct in-depth follow-up observations, thereby potentially identifying host galaxies. The above facts highlight that our results can be considered a conservative analysis.}

Our results show that the dark sirens of the Taiji-TianQin-LISA could also play a crucial role in helping solve the Hubble tension. It is worth expecting that the Hubble tension could be solved with the help of the Taiji-TianQin-LISA network.

\section{Conclusion}\label{sec:con}

In this work, we show the potential of the future standard sirens from the Taiji-TianQin-LISA network in constraining cosmological parameters. We consider the optimistic and conservative scenarios of the standard siren observations, i.e., the bright and dark sirens. Three population models of MBHB, i.e., pop III, Q3d, and Q3nod, are used to simulate the MBHB catalogs.
For the simulations of bright sirens, we calculate the expected detection rates of the GW-EM events based on the three MBHB models. For comparison, we also show the results of the single TianQin observatory and the Taiji-TianQin network. For the analysis of the simulated bright siren data on constraining cosmological parameters, we choose three typical cosmological models, i.e., the $\Lambda$CDM, $w$CDM, and $w_0w_a$CDM models.
In the dark siren case, we used the simulated GW events at $z<3$ combined with the simulated galaxy catalog to estimate cosmological parameters in the $\Lambda$CDM model.

We find that the bright siren data from the Taiji-TianQin-LISA network based on the pop III and Q3nod models could give tight constraints on the Hubble constant, with both constraint precisions being 0.9\%. Nevertheless, the EoS parameters of dark energy could not be well constrained by the bright siren data from the Taiji-TianQin-LISA network (pop III and Q3nod). Fortunately, the Taiji-TianQin-LISA network and CMB show different parameter degeneracy orientations, especially in the dynamical dark energy models. Therefore, the combination of the Taiji-TianQin-LISA network and CMB could break cosmological parameter degeneracies and improve the constraint precisions of cosmological parameters. In addition, the Taiji-TianQin network gives comparable constraint results with those of the Taiji-TianQin-LISA network, which may indicate that the two space-based detector network is sufficient to constrain cosmological parameters well. Since the expected detection number of GW-EM events based on the Q3d model is small, Q3d gives loose constraints on all cosmological parameters. Concretely, the constraint precisions of $\Omega_{\rm m}$ and $H_0$ using CMB+Taiji-TianQin-LISA (pop III) are 1.7\% and 0.6\%, which are close to or better than 1\% (the standard of precision cosmology). While in the $w$CDM model, CMB+Taiji-TianQin-LISA (pop III) gives $\sigma(w)=0.036$, which is 85.6\% better than the constraint given by the CMB data. In the $w_0w_a$CDM model, CMB+Taiji-TianQin-LISA (pop III) gives $\sigma(w_0)=0.23$ and $\sigma(w_a)=0.63$. For the dark sirens from the Taiji-TianQin-LISA network, the constraint precisions of $H_0$ based on the pop III, Q3d, and Q3nod models are {1.29\%, 8.86\%, and 1.17\%}, respectively, indicating that the dark sirens from the Taiji-TianQin-LISA network may also play a crucial role in measuring the Hubble constant. {In addition, compared to the Taiji-TianQin network and the Taiji-LISA network, the Taiji-TianQin-LISA network can improve the constraints on the Hubble constant by 51.9\% and 28.9\%.}

Hence, we can conclude that: (i) the bright sirens from the Taiji-TianQin-LISA network could precisely measure the Hubble constant with a precision of 0.9\%, but are poor at measuring dark energy; (ii) the bright sirens from the Taiji-TianQin-LISA network could effectively break the cosmological parameter degeneracies generated by the CMB data, especially in the dynamical dark energy models; (iii) the dark sirens from the Taiji-TianQin-LISA network could also provide a tight constraint on the Hubble constant, with a precision of 1.2\%. It is worth expecting to use the Taiji-TianQin-LISA network to help solve the Hubble tension and explore the nature of dark energy.

\begin{acknowledgments}
This work was supported by the National SKA Program of China (Grant Nos. 2022SKA0110200 and 2022SKA0110203), the National Natural Science Foundation of China (Grant Nos. 11975072, 11875102, and 11835009), the National 111 Project (Grant No. B16009), and the Fundamental Research Funds for the Central Universities (Grant No. N232410019). We thank Liang-Gui Zhu and Yue Shao for helpful discussions.
\end{acknowledgments}

\appendix
\renewcommand\thefigure{\Alph{section}\arabic{figure}}
\renewcommand\thetable{\Alph{section}\arabic{table}}

\section*{Appendix}

\section{Warning time of EM counterparts}\label{Appendix:warning time}

\begin{figure}[!htbp]
\includegraphics[width=0.45\textwidth]{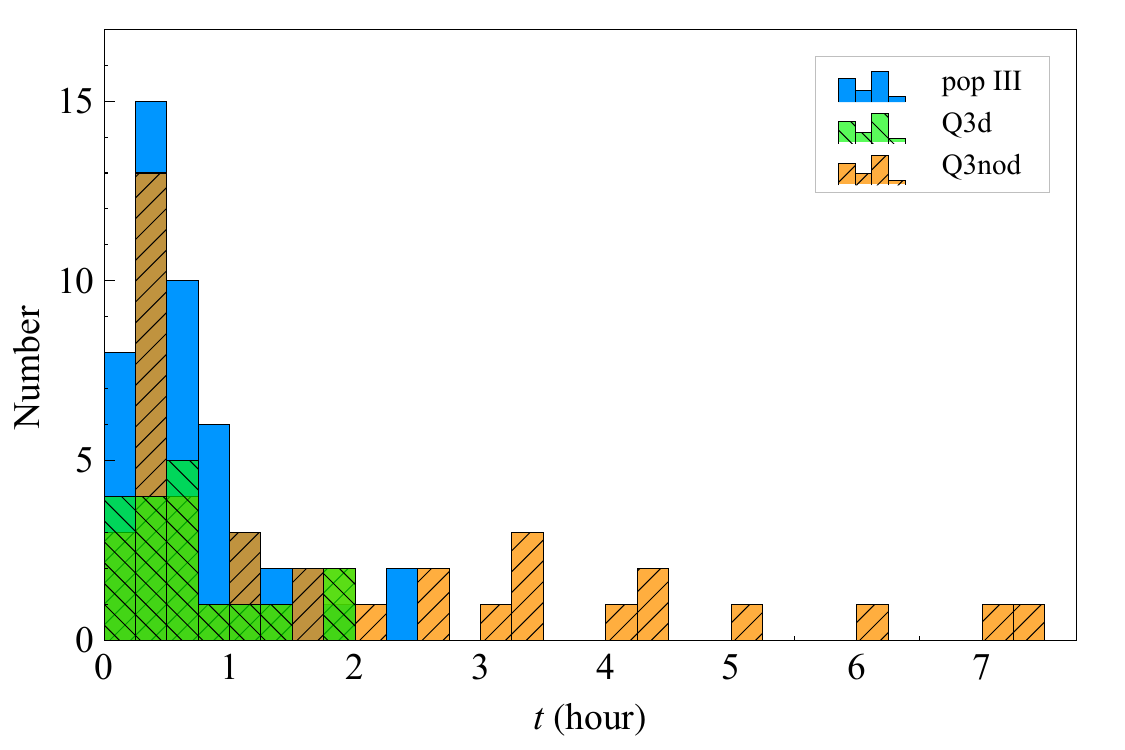}
\centering
\caption{\label{fig9} The warning time distributions of the simulated bright sirens based on the pop III, Q3d, and Q3nod models.}
\end{figure}

In this Appendix, we provide more information about the detections of EM counterparts. In this work, we adopt the inspiral-only GW waveform and simulate the bright sirens. It is commonly believed that the EM radiations are emitted in the mergers of MBHBs. Hence, the warning times of EM counterparts are important for the detections of EM counterparts.
The warning times of EM counterparts $t_{\mathrm{warning}}$ can be calculated by the integration upper limit $f_{\rm upper}$. We have
\begin{equation}
t_{\mathrm{warning}}=\frac{5\mathcal{M}_{\rm chirp}}{{(8\pi\mathcal{M}_{\rm chirp}f)}^{8/3}}.
\end{equation}
Here we use $f=f_{\rm upper}$ to calculate the warning time. Note again that we assume the real-time data analysis of the GW source.
In actual observation, $f$ should be replaced by the frequency that meets $\rho\geq 8$ and $\Delta \Omega\leq 10\ \mathrm{deg}^2$. In our analysis, we consider that the warning could only be issued once $f$ reaches $f_{\rm upper}$. In Fig.~\ref{fig9}, we show the warning time distributions of the simulated bright sirens based on the pop III, Q3d, and Q3nod models. We can see that the warning times are on the order of hours.

\section{Constraint results of cosmological parameters}\label{Appendix:cosmological parameter}

\begin{figure}[!htb]
\begin{center}
\includegraphics[width=\linewidth,angle=0]{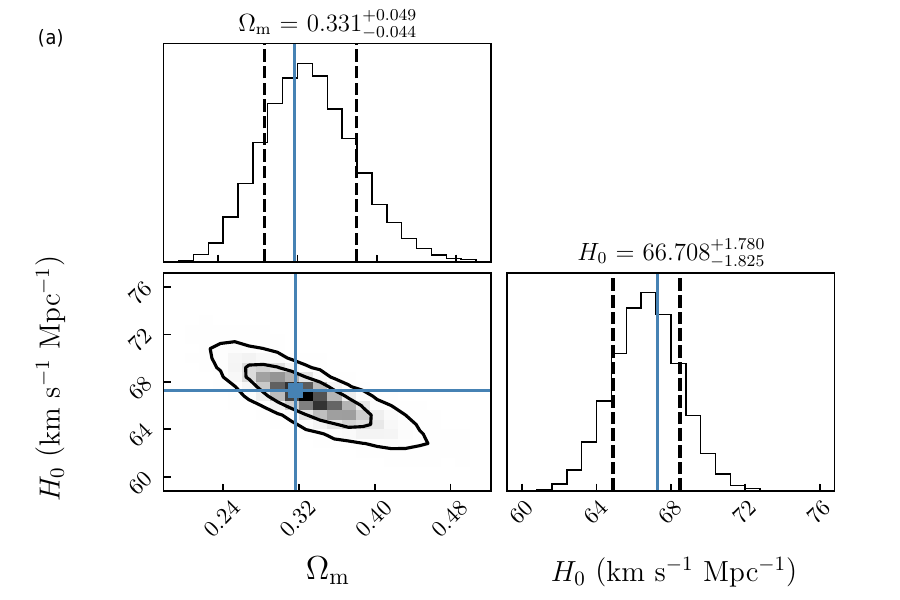}\label{Fig10a} \hspace*{.0cm}
\includegraphics[width=\linewidth,angle=0]{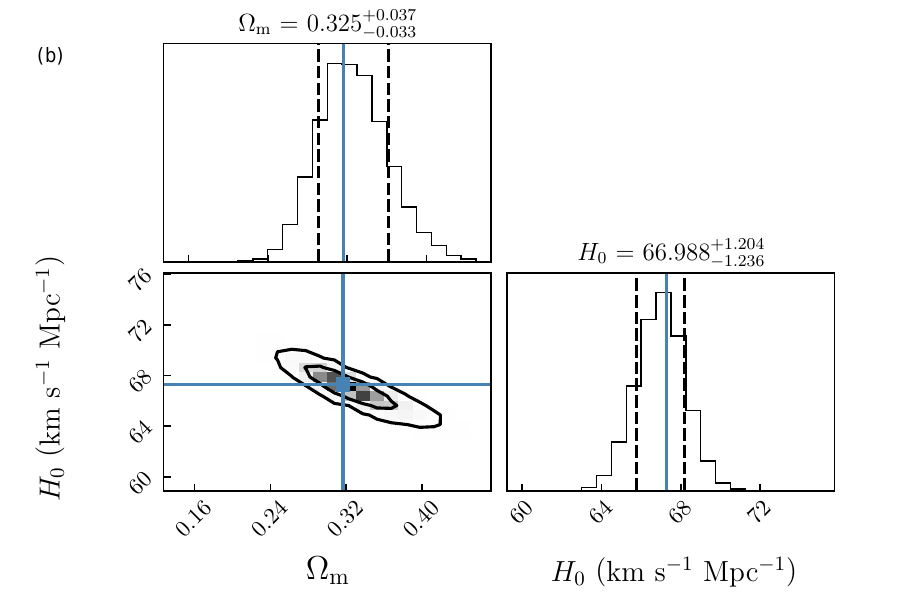}\label{Fig10b}\\
\includegraphics[width=\linewidth,angle=0]{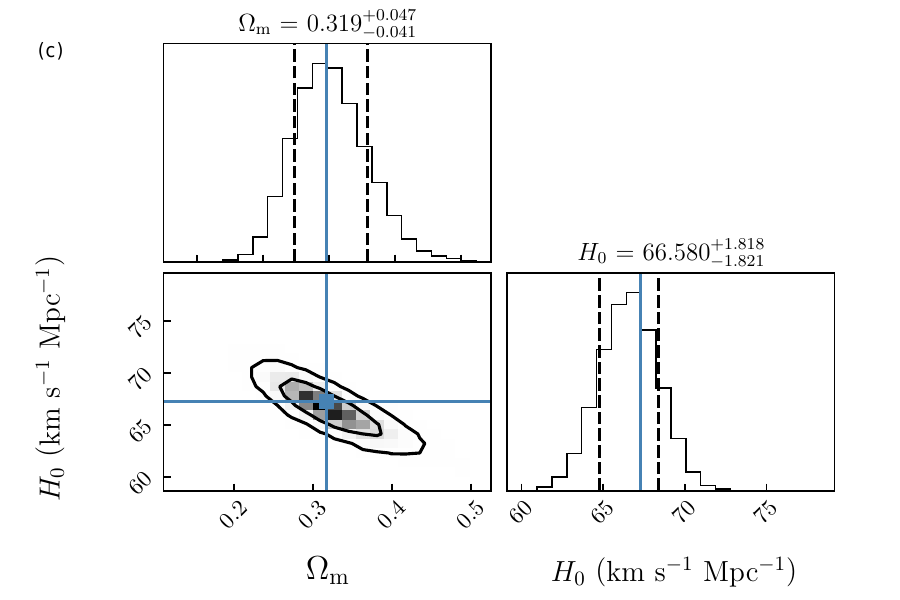}\label{Fig10c}
\end{center}
\caption{Same as Fig.~\ref{fig8}, but for the Taiji-TianQin network (a), the Taiji-LISA network (b), and the TianQin-LISA network (c) based on the pop III model.}\label{fig10}
\end{figure}

In this Appendix, we show the detailed constraint results of bright sirens and dark sirens. In Table~\ref{tab5}, we show the constraint results in the $\Lambda$CDM, $w$CDM, and $w_0w_a$CDM models using the simulated bright siren data from TianQin, the Taiji-TianQin network, and the Taiji-TianQin-LISA network based on the pop III, Q3d, and Q3nod models. In Fig.~\ref{fig10}, we show the constraints on the $\Lambda$CDM model using the mock dark siren data from the Taiji-TianQin network, the Taiji-LISA network, and the TianQin-LISA network based on the pop III model.

\begin{table*}[!htbp]
\renewcommand{\arraystretch}{1.5}
\caption{The absolute errors ($1\sigma$) and the relative errors of the cosmological parameters in the $\Lambda$CDM, $w$CDM, and $w_0w_a$CDM models using the simulated bright siren data from TianQin, the Taiji-TianQin network, and the Taiji-TianQin-LISA network based on the pop III, Q3d, and Q3nod models. Note that $H_0$ is in units of ${\rm km\ s^{-1}\ Mpc^{-1}}$.}
\label{tab5}
\begin{center}{\centerline{
\begin{tabular}{|m{1.7cm}<{\centering}|m{2.9cm}<{\centering}|m{1.4cm}<{\centering}|m{1.1cm}<{\centering}|m{1.1cm}<{\centering}|m{1.1cm}<{\centering}|m{1.1cm}<{\centering}|m{1.1cm}<{\centering}|m{1.1cm}<{\centering}|m{1.1cm}
<{\centering}|m{1.1cm}<{\centering}|m{1.1cm}<{\centering}|m{1.1cm}<{\centering}|}
\hline
       Model &Detection strategy& MBHB model & {$\sigma(\Omega_{\rm m})$} & {$\sigma(H_0)$}&{$\sigma(w)$}  &{$\sigma(w_0)$} &{$\sigma(w_a)$}   &{$\varepsilon(\Omega_{\rm m}$)}&{$\varepsilon(H_0)$}&{$\varepsilon(w)$}&{$\varepsilon(w_0)$}\\
            \hline
                         & \text{} &pop III &$0.041$ & $1.10$  & $-$& $-$& $-$& $12.5\%$&$1.6\%$ &$-$ & $-$ \\
\multirow{6}{*}{$\Lambda$CDM}  &TianQin &Q3d &$0.245$  & $9.95$  &$-$ &$-$ &$-$ & $52.1\%$&$15.7\%$ &$-$ &  $-$ \\
                       & \text{} &Q3nod&$0.057$  & $1.20$  &$-$ &$-$ &$-$ & $17.1\%$&$1.8\%$ & $-$&$-$  \\
                       \cline{2-12}
                       & \text{} &pop III &$0.019$  & $0.63$  &$-$ &$-$ &$-$ & $6.0\%$&$0.9\%$ &$-$ & $-$ \\
                      &Taiji-TianQin &Q3d&$0.106$  & $5.60$  & $-$& $-$& $-$& $28.4\%$&$8.5\%$ &$-$ &$-$   \\
                       & \text{} &Q3nod&$0.021$  & $0.64$  &$-$ &$-$ &$-$ & $6.6\%$&$1.0\%$ & $-$&$-$  \\
                       \cline{2-12}
                      & \text{} &pop III &$0.019$  & $0.63$  &$-$ &$-$ &$-$ & $5.8\%$&$0.9\%$ &$-$ & $-$ \\
                      &Taiji-TianQin-LISA &Q3d&$0.096$  & $5.15$  & $-$& $-$& $-$& $26.1\%$&$7.9\%$ &$-$ &$-$   \\
                       & \text{} &Q3nod&$0.019$  & $0.62$  &$-$ &$-$ &$-$ & $5.8\%$ & $0.9\%$ & $-$&$-$ \\\hline
                     & \text{} &pop III &$0.057$  & $2.60$  &$0.395$ &$-$ & $-$& $17.8\%$&$3.8\%$ &$35.0\%$ &$-$ \\
\multirow{6}{*}{$w$CDM} &TianQin &Q3d&$0.275$  & $9.90$  & $0.695$&$-$ &$-$ & $56.1\%$&$16.3\%$ &$73.9\%$ &$-$   \\
                      & \text{} &Q3nod&$0.077$  & $2.90$  &$0.520$ &$-$ &$-$& $24.7\%$&$4.3\%$ &$44.8\%$ &$-$ \\
                      \cline{2-12}
                      & \text{} &pop III &$0.026$  & $1.50$  &$0.195$ &$-$ &$-$ & $8.0\%$&$2.2\%$ &$18.4\%$ &$-$ \\
                      &Taiji-TianQin &Q3d&$0.116$  & $9.95$  & $0.650$& $-$&$-$ & $34.2\%$&$14.9\%$ &$62.5\%$ &$-$   \\
                      & \text{} &Q3nod&$0.030$  & $1.45$  &$0.205$ &$-$ & $-$& $9.2\%$&$2.1\%$ &$19.3\%$ &$-$ \\
                      \cline{2-12}
                      & \text{} &pop III &$0.026$  & $1.45$  &$0.190$ &$-$ &$-$ & $8.0\%$&$2.1\%$ &$17.9\%$ &$-$ \\
                      &Taiji-TianQin-LISA &Q3d&$0.105$  & $9.80$  & $0.630$& $-$&$-$ & $32.0\%$&$14.7\%$ &$60.6\%$ &$-$   \\
                      & \text{} &Q3nod&$0.027$  & $1.40$  &$0.200$ &$-$ & $-$& $8.5\%$&$2.1\%$ &$18.9\%$ &$-$ \\
                      \hline
                         & \text{} &pop III &$0.063$  & $4.15$  &$-$&$0.680$ & $2.30$& $18.4\%$&$6.0\%$ & $-$&$57.6\%$\\
\multirow{6}{*}{$w_0w_a$CDM} &TianQin &Q3d&$0.260$  & $14.00$  &$-$ & $-$&$-$ & $50.0\%$&$22.6\%$ &$-$ &$-$  \\
                      & \text{} &Q3nod&$0.081$  & $5.30$  &$-$ &$-$ & $-$ & $24.0\%$& $7.5\%$ &$-$ & $-$ \\
                 \cline{2-12}
                       & \text{}&pop III  &$0.053$  & $2.60$  &$-$ &$0.420$ & $2.05$& $15.9\%$&$3.9\%$ & $-$&$50.0\%$\\
                       &Taiji-TianQin &Q3d&$0.150$  & $10.00$  & $-$&1.130 & $-$& $38.9\%$&$14.9\%$ &$-$ &$-$  \\
                      & \text{} &Q3nod&$0.056$  & $2.10$ &$-$ &$0.340$ &$1.80$ &  $17.4\%$&$3.1\%$ & $-$&$37.3\%$ \\
                      \cline{2-12}
                       & \text{}&pop III  &$0.052$  & $2.60$  &$-$ &$0.420$ & $2.00$& $15.6\%$&$3.9\%$ & $-$&$50.0\%$\\
                       &Taiji-TianQin-LISA &Q3d&$0.140$  & $10.00$  & $-$&$1.120$ & $-$& $37.8\%$&$14.7\%$ &$-$ &$-$  \\
                      & \text{} &Q3nod&$0.056$  & $2.00$ &$-$ &$0.330$ &$1.70$ & $15.5\%$ & $3.0\%$ & $-$& $36.7\%$ \\
                     \hline
\end{tabular}}}
\end{center}
\end{table*}

\bibliography{ttl_network}
\end{document}